 \documentclass[final,3p,times,twocolumn,authoryear]{elsarticle}

\usepackage{amssymb}
\usepackage{xcolor}
\usepackage{ulem}
\usepackage{booktabs}
\usepackage[flushleft]{threeparttable}
\usepackage{rotating}
\usepackage{lscape}
\usepackage{hyperref}
\usepackage{fancyhdr} 
\pdfoutput=1 

\def\fm{\hbox{$.\!\!^{\rm m}$}}

\def\cd{\hbox{\,d$^{-1}$\,}}
\newcommand{\bfr}[1]{\textcolor{red}{\textbf{#1}}}

\journal{New Astronomy}

\begin{document}

\begin{frontmatter}



\title{Unveiling $\delta$ Scuti and $\gamma$ Doradus Hybrid Pulsation of HD 53166 and HD 53349 Plus Rich Frequencies in HD 52788 -- Catalogs of 70552 $\delta$ Sct and 8080 $\gamma$ Dor\tnoteref{nt0} } 
\tnotetext[nt0]{This research is dedicated to my wife Jingyun Zhang who has been supporting my works all the time.  }

\fancyhead[L]{} 
\fancyhead[C]{to appear in New Astronomy, Volume 105, January 2024} 
\fancyhead[R]{} 

\author{Ai-Ying Zhou\corref{cor1}}
\ead{aiying@nao.cas.cn}
\cortext[cor1]{To appear in New Astronomy, Volume 105, January 2024 
}
\affiliation{organization={National Astronomy Observatories, Chinese Academy of Sciences},
            addressline={A20 Datun Road, Chaoyang District}, 
            city={Beijing},
            postcode={100101}, 
            country={P.R. China}}

\begin{abstract}    
We present discoveries of the hybrid $\delta$ Scuti and $\gamma$ Doradus types stellar pulsational variability of two stars, 
HD 53166 and HD 53349, which were previously used as comparison in the differential photometry of a known $\delta$ Scuti star HD 52788. 
More surprisingly, HD 52788 exhibits 126 pulsation frequencies in a small domain in 4--12\,d$^{-1}$ based on 24 sectors light curves of \textit{TESS}. 
We compared the three stars with up-to-date class members collected from existing catalogs in multiple 
Hertzsprung-Russell diagrams for checking group property. 
Two comprehensive catalogs of 70\,552 $\delta$ Sct and 8\,080 $\gamma$ Dor stars have been cross-identified with the \textit{TESS} Input Catalog (TIC) and \textit{Gaia} DR3. 
The catalogs include multiple stellar parameters extracted from both \textit{Gaia} DR3 and TIC v8.2. 
The rich pulsation contents turn HD 52788 into a rare distinctive object 
in its class for testing current stellar evolution and pulsation models.    
\end{abstract}




\end{frontmatter}


\section{Introduction}
\label{sect:intro}

The $\delta$ Scuti and $\gamma$ Doradus stars are both pulsating variables that are situated on and near the region of the A and F-type main sequence, inside the lower part of classical Cepheid instability 
strip~\citep{2009AIPC.1170..403H, 2001A&A...366..178R,2000ASPC..210....3B}.
 
$\delta$ Scuti stars (DSCT) are short-period ($\lesssim$0.3 day) pulsators situated inside the lower part of 
the classical Cepheid instability strip on and near the main sequence (MS) in the Hertzsprung--Russell diagram, 
with spectral types ranging from about A2 to F6 for luminosity classes III (giants), IV (subgiants), and V (dwarfs, main sequence stars)~\citep{2018MNRAS.475..478Q, 2014CoSka..43..338S, 2009ssc..book.....G}.
 They pulsate in radial as well as non-radial acoustic modes excited mainly in the $\kappa$ mechanism~\citep{2000ASPC..210....3B, 
2016MNRAS.457.3163X}.
The majority of $\delta$ Sct stars are multiple mode pulsators.
Most of the $\delta$ Sct stars belong to the Population I stars that cover a mass range from about 
1.5 to 2.3\,M$_{\odot}$~\citep{2019MNRAS.485.2380M}.

$\gamma$ Doradus-type variables (GDOR) are a relatively new and small young group of pulsators 
found overlap or just beyond the lower red (cool) edge of the $\delta$ Scuti instability strip.
This class of variable stars is first characterized in late 1990s~\citep{1995IBVS.4195....1K,1999PASP..111..840K, 1999MNRAS.309L..19H}, 
while $\gamma$ Doradus itself as the prototype of its class was first revealed to be variable by~\citet{1963MNSSA..22...65C}. 
These stars display longer periods and 
low-amplitude variability that was confirmed to be caused or characterized by
high-order ($n$), low-degree ($\ell$), non-radial, photospheric gravity-type oscillations ($g$-mode pulsations, 
\citealt{1994MNRAS.270..905B, 1999PASP..111..840K}) 
rather than the $p$-mode pulsations of most $\delta$ Scuti variables
~\citep{1993AAS...183.8402K,1995IBVS.4195....1K,1998IAUS..185..339K,1999PASP..111..840K,1999MNRAS.309L..19H}.
Their multiple periods spaced from several tenths of a day to slightly in excess of one day and pulsational amplitudes 
usually do not exceed 0.1\,mag.
Most GDOR stars are typically early F or late A type main sequence stars with ages of $\sim$150\,Myr.  
There are hundreds of bona fide GDOR memebers and a few thousand candiates by far~\citep{2022AJ....163..180H, 2020MNRAS.493.5871B, 2022arXiv221210776B}.  


In recent years, a number of stars have been found that exhibit both $\delta$ Scuti and $\gamma$ Doradus type pulsations, such as those discovered in ground observations~\citep{2005AJ....129.2026H, 2009MNRAS.398.1339H, 2010ApJ...713L.192G, 2011ApJ...743..153H, 2012ASPC..462..111H, 2012AJ....144..141Z, 2017MNRAS.466..122K} 
as well as from space missions~\citep{2015MNRAS.452.3073B, 2019MNRAS.490.4040A, 2020ApJ...895..124Z, 2020MNRAS.493.5871B} 
and even hybrid pulsators in binary systems~\citep{2019ApJ...884..165Z, 2022ApJ...937...80M}.
Hybrid pulsators, generally speaking, are pulsating stars that show two different types of oscillation features or classified with multiple identities~\citep{2015arXiv150105166Z}. 
Particularly in this paper, it refers to the stars that show both $\gamma$ Doradus type $g$ modes and $\delta$ Scuti type $p$ modes. 
The major observational problem involved in a hybrid pulsator that needs to be solved before arriving at a unique seismic model 
is more severe than single oscillating type stars -- a sufficiently large number of pulsation modes must be detected and identified 
in both DSCT and GDOR pulsation frequency domains~\citep{2009CoAst.159...42H}.

On the side of oscillation theory, the low-frequency $g$ modes in $\gamma$ Dor stars are excited by convective flux modulation~\citep{2000ApJ...542L..57G},
but it is not known whether a single excitation mechanism might be responsible for both the $p$ and $g$ modes in hybrid stars~\citep{2015MNRAS.452.3073B}.
A recent theoretic study shows that 
oscillations of $\delta$ Scuti and $\gamma$ Doradus stars are both due to the combination of $\kappa$ mechanism and 
the coupling between convection and oscillations, and they belong to the same class of variables at the low-luminosity part 
of the Cepheid instability strip. Within the $\delta$ Sct-$\gamma$ Dor instability strip, 
most of the pulsating variables are very likely hybrids that are excited in both $p$ and $g$ modes~\citep{2016MNRAS.457.3163X}.

The $\gamma$ Dor, $\delta$ Sct, especially those with rich spectrum of pulsation frequencies and the hybrid
$\delta$ Sct/$\gamma$ Dor pulsators offer tremendous potential for asteroseismology~\citep{2009AIPC.1170..455P, 2019MNRAS.490.4040A, 2021RvMP...93a5001A, 2022AJ....163..180H}.  
Thus the hybrid pulsators are promising targets for asteroseismology and have the potential to advance the theories of both stellar evolution and pulsation. 
It is widely recognized to be worthy of detection and characterization of new hybrid pulsators, as their asteroseismic studies 
would provide new insights into the physics of the oscillations~\citep{2022AJ....163..180H, 2010ApJ...713L.192G}.
Therefore, the detection of hybrid pulsators is astrophysically signifcant.

Another interesting research related to DSCT and GDOR is the detection of DSCT or GDOR in eclipsing binary systems, 
such as those recent works by~\cite{2013A&A...556A..87C,2020MNRAS.497.4363L, 2022ApJS..263...34C, 2022MNRAS.515.2755S}.

HD 52788 
was announced to be a $\delta$ Scuti star by \citet{kurtz1979}.
Based on a total of 104 hours of differential photometric observations obtained in 1978--1980, 
both \citet{kurtz1981} and \citet{zhou2004} failed to establish a consistent frequency solution for the pulsational behaviour of the star.
The pulsation frequency spectrum of HD 52788 changed yearly with somewhat complex light variability,  
makes it a special target in its class.

\begin{figure}[thb]
   \vspace{0.5mm}
   \hspace{-13mm}\includegraphics[width=75mm, height=105mm,angle=-90,scale=0.99]{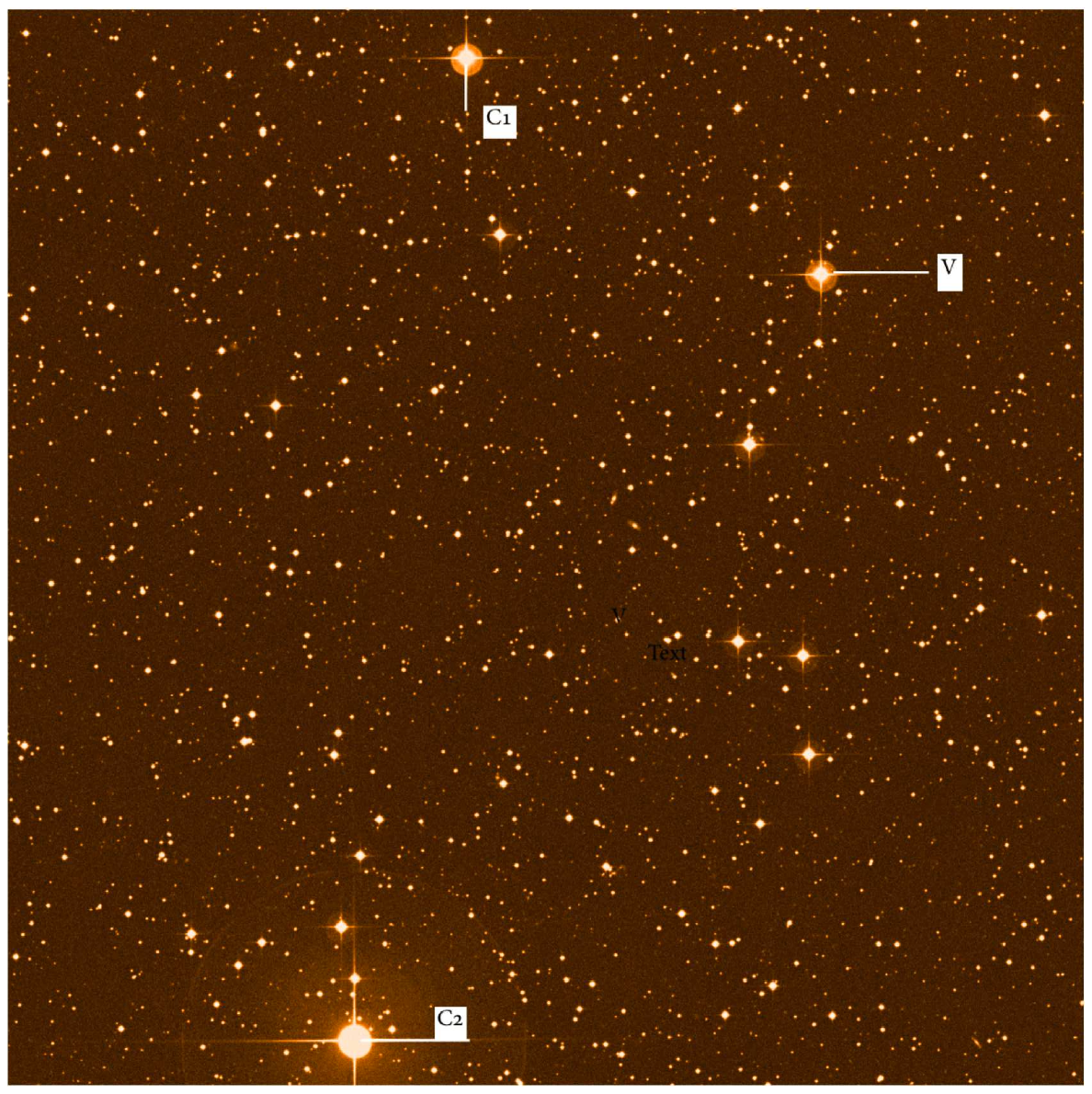}
   \caption[]{Finder chart of HD 52788 and comparisons used in Kurtz(1979,1981).  }
   \label{fig:chart}
\end{figure}

What makes the author initially interested is the cause of the unstable pulsation frequency spectrum of HD 52788. 
The poor frequency resolution could result from inadequate time series data, 
intrinsic rapid changing, and data issues using wrong stars as comparison among others.   
Which is the true source? 
Potential comparison stars in the target field were often discovered to be new variables, for example, 
\citet{du1999, zhou2006, jurc2012}. 
Due to the star in the southern sky, while the interested observers in the northern hemisphere, the question is that could we have a chance to obtain adequate data to derive a unique set of frequencies that represents the light variations of this star. 
Fortunately, the  NASA's Transiting Exoplanet Survey Satellite (\textit{TESS},~\citealt{TESS}) data 
provide us with the best opportunity to check any star's light variations more precisely.
We are delighted with good fortune when checking \textit{TESS} data.
We found rich intrinsic pulsation contents of HD 52788 based on the perfect data sets, 
and serendipitously we reveal that 
the two comparison stars used in Kurtz's differential photometry turned out to be pulsating variable stars (see Fig.\,\ref{fig:chart}). 
This fact of using the wrong comparison would have added uncertainties in resolving the target's pulsation. 

This work analyzes the \textit{TESS} light curves of three stars HD 52788 and the two previously used nearby comparison stars HD 53166 and HD 53349.
In Section 2, we describe the \textit{TESS} data and our custom processings, followed by pulsational analyses of the three stars in Section 3. 
Results and discussions are summarized in the last two sections.  
 
\section{\textit{TESS} Data}
 
\textit{TESS} is an MIT-led NASA mission dedicated to discovering transiting exoplanets orbiting nearby bright stars 
by an all-sky photometry survey~\citep{TESS}. \textit{TESS} spacecraft launched on April 18, 2018 and its regular science operations began on July 25, 2018.
\textit{TESS} rotates every $\sim$13.7 days per cycle along a unique highly elliptical lunar-resonant orbit around the Earth (about 600\,km from Earth).  
\textit{TESS} is equipped with four identical refractive cameras with a combined field-of-view (FOV) of 24$\times$96 degrees (a segment of sky, 
known as an observing sector). 
The lens assembly in each camera has a 10.5\,cm diameter entrance pupil (aperture) and a focal ratio f/1.4.
Each of the four cameras has four 2048$\times$2048 CCD detectors (i.e. total 16 CCDs). 
Each detector pixel sized 15$\times$15 micron corresponds to 21 arcseconds in the sky.
\textit{TESS} collects light in 600--1000\,nm centered on the traditional Cousins $I$-band ($I_{\rm C}$, central wavelength = 786.5\,nm). 
Please refer to the details to \textit{TESS} Science Data Products Description 
Document\footnote{https://archive.stsci.edu/missions-and-data/tess;~ 
https://archive.\\stsci.edu/files/live/sites/mast/files/home/missions-and-data/active-\\missions/tess/\_documents/EXP-TESS-ARC-ICD-TM-0014-Rev-F.pdf} and 
``Characteristics of the \textit{TESS} space telescope" web page\footnote{https://tess.mit.edu/science/ or https://heasarc.gsfc.nasa.gov/docs/\\tess/the-tess-space-telescope.html}. 
\textit{TESS} filter basically matches with \textit{Gaia} Red Photometer with bandpass spanning in 630--1050\,nm~\citep{GaiaMission}.


\textit{TESS} cameras actually expose at a cadence of 2 seconds. That is, the CCDs take images and read them out continuously at 2-second intervals. 
However, the 2-second frames are used for spacecraft guiding, they were not downloaded to the ground. 
The images are processed on the spacecraft by the data handling unit (DHU, a Space Micro Image Processing Computer). 
The DHU stacks 2-second images in groups of 60 or 900 to produce 2-minute cadence images (in cut Target Pixel File, TPF) 
and 30-minute cadence for general observations.
High cadence is needed for the detection of exoplanets, so exposures of planet search targets and 
other stars of particular interest (cataloged 200,000 primary stars) are obtained every 2 minutes 
while the Full-Frame Images (FFIs) of the entire field of view are returned every 30 minutes.
Finally, pixels in postage stamps around \textit{TESS} mission target stars will be downloaded at a 2-minute cadence, 
while FFIs will be downloaded at the 30-minute cadence. 
These two sets of data will allow general variability studies for the vast majority of stars.
The data on spacecraft are transmitted to Earth when the spacecraft reaches orbital perigee every $\sim$13.7 days.
Each sector of the sky will be observed twice with 27.4 days observing period. 
So the \textit{TESS} observation is most sensitive to exoplanets with periods of less than 13 days 
so that at least two transits are used for discovery. 

In the meantime, \textit{TESS} high-precision uninterrupted photometry and time resolution (2-minute and 30-minute cadence)  
are perfect for stellar pulsation and asteroseismology 
(e.g. the works by~\citealt{2021ApJS..257...53L, 2021AJ....162..170H}).
Furthermore, \textit{TESS} produces additional pixels in little postage stamps surrounding a few bright asteroseismology targets  and downloads at 
20-second cadence (1000 stars, e.g. the works by \citealt{2022AJ....163...79H}).
After \textit{TESS} first two-year prime mission, started in July 2020, \textit{TESS} was revisiting the sky in 
an extended ongoing mission that records full-frame images at a fast ten-minute cadence. 
\textit{TESS} provides a golden opportunity for studying short to moderate long periodic variables, 
in special, $\delta$ Sct and $\gamma$ Dor stars~\citep{2019MNRAS.490.4040A}, RR Lyr stars and so on.

We first checked the \textit{TESS} archive at the MAST Portal (Mikulski Archive for Space Telescopes\footnote{https://mast.stsci.edu/portal/Mashup/Clients/Mast/Portal.html; or  https://archive.stsci.edu/tess/}) 
for available data of the interested target star  
HD 52788 (=V383 Carinae =TIC 279361762 = ASASSN-V J065904.08-583054.1, 
$\alpha_{2000}$=06:59:04, $\delta_{2000}$=$-$58:30:53 [104.7667$^{\circ}$, $-58.5147^{\circ}$], $V$=8\fm40) 
and the two comparison stars used in literature C1=HD 53166 (=TIC 279431011, $V$=8\fm1) 
and  C2=HD 53349 (=TIC 279476396, $V$=6\fm01). 
We downloaded the NASA's Science Processing Operations Centre (SPOC) generated files, 
including both the extracted light curves (*-s\_lc.fits) and Target Pixel files (*-s\_tp.fits). 
The light curve files *-s\_lc.fits will be read directly with {\sc astropy} and {\sc lightkurve} packages~\citep{astropy,lightkurve} 
in a Python script following the \textit{TESS} Archive Manual\footnote{
The Beginner Tutorial Notebooks: https://outerspace.stsci.edu/\\display/TESS/TESS+Archive+Manual;
https://github.com/spacetelescope/notebooks/blob/master/notebooks/\\MAST/TESS/beginner\_how\_to\_use\_lc/beginner\_how\_to\_use\_lc.ipynb}.
The \textit{TESS} data information for these three stars is listed in Table~\ref{tab:TESS-data}.

\subsection{SAP and PDCSAP}
\begin{figure}[thb]
   \vspace{1mm}
   \hspace*{0.9mm}\includegraphics[width=0.45\textwidth, height=30mm,angle=0,scale=0.99]{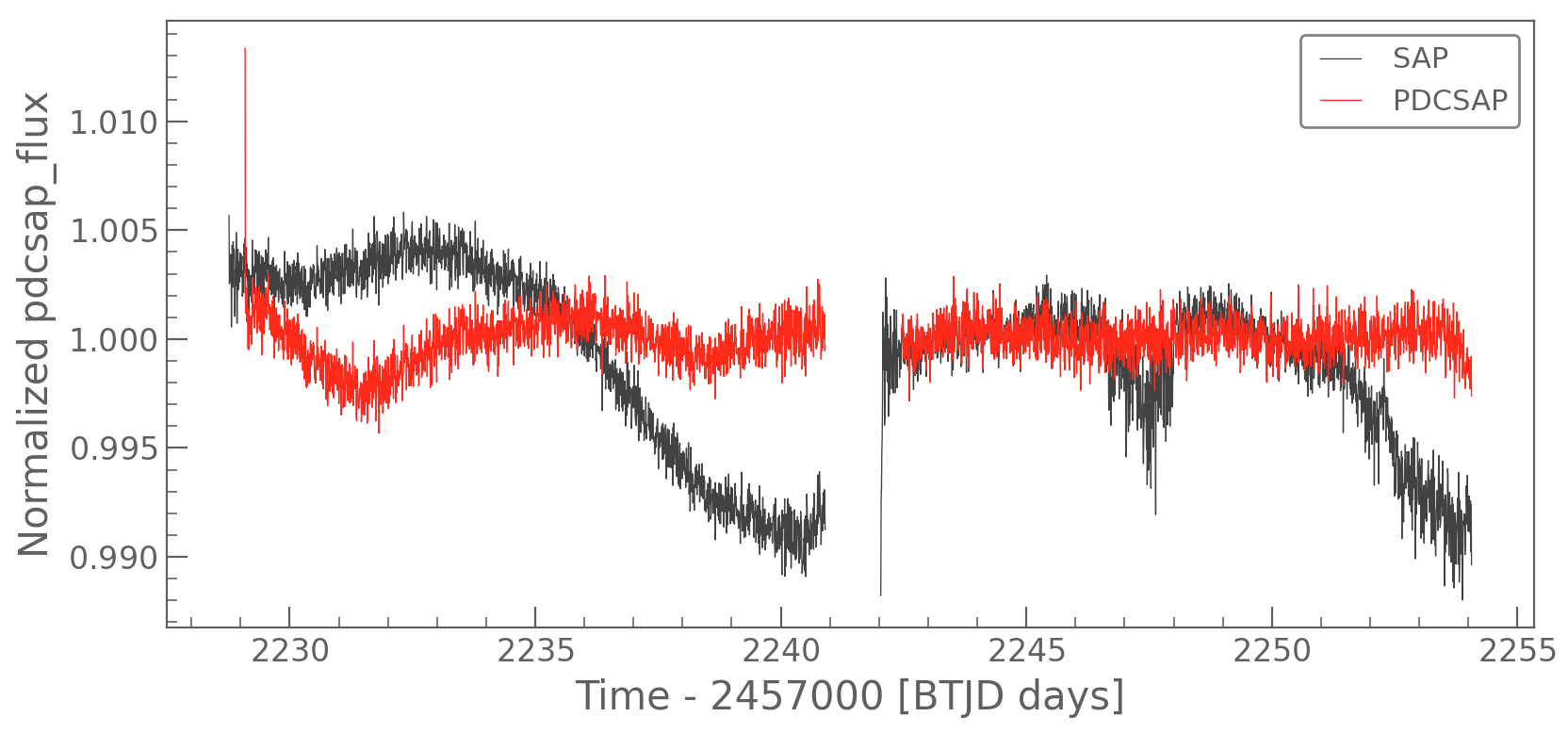}
   \hspace*{0.9mm}\includegraphics[width=0.45\textwidth, height=30mm,angle=0,scale=0.99]{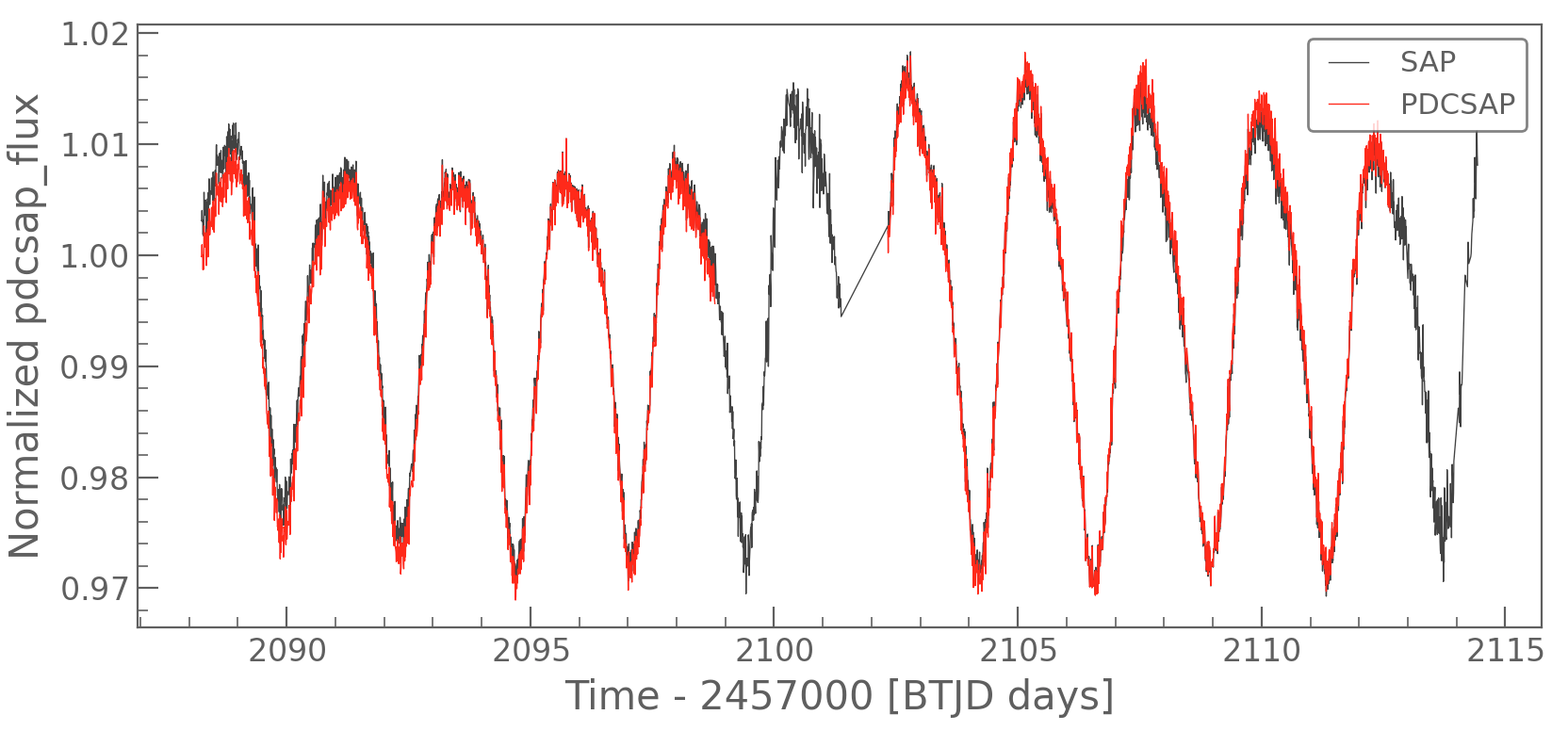}
   \caption[]{Comparison between TESS SAP and PDCSAP fluxes of two variable stars. Up: this case PDCSAP will be used; 
down: SAP can be used. } 
   \label{fig:SAP-PDCSAP}
\end{figure}

There are two kinds of SPOC-extracted light curves:
one is Simple Aperture Photometry flux (SAP), which is the flux after summing the calibrated pixels within 
the \textit{TESS} optimal photometric aperture, 
and the other is Pre-search Data Conditioned Simple Aperture Photometry (PDCSAP), 
it is the SAP flux from which long-term trends have been removed using so-called Co-trending Basis Vectors  
and nominally corrected for instrumental variations and excessive scattered light removed. 
Figure~\ref{fig:SAP-PDCSAP} shows two cases of the two kinds of light curves.
PDCSAP flux is usually cleaner data than the SAP flux and will have fewer systematic trends.
Thus PDCSAP flux is widely used for final analysis without further processing (e.g. \citealt{2019A&A...627A..43D, 
2020A&A...636A..58A, 2020A&A...642A..49D, 2021MNRAS.503.4092B}).  
However, the PDCSAP flux might suffer from loss or wrong-deletion of long-term and transient burst variability 
intrinsic to stars, for instance, as that pointed out by~\citealt{hill2022, litt2021}.  
With these concerns, SAP flux is used by some authors (e.g.~\citealt{essen2020, stei2021, 2021ApJ...919..131H, 
2022ApJS..258...16P, 2022arXiv220709169S}) accompanied with additional custom processing 
such as detrendings depending on science goals. 

We then checked the two comparison stars in General Catalog of Variable Stars (GCVS database, Version 2022 Jun. 
and 5.1,~\citealt{2017ARep...61...80S})\footnote{online URL: http://www.sai.msu.su/gcvs/cgi-bin/search.htm\#cat} and 
Revised Version of the New Catalogue of Suspected Variable Stars (NSV Release 2, 1982, 1998,~\citealt{2022ARep...66..555K})\footnote{https://heasarc.gsfc.nasa.gov/W3Browse/all/gcvsnsvars.html}, 
and found that C2=HD 53349 =HR 2661 =NSV 3349 is a dwarf (DM, spectral type F0V) and a new suspected variable star but without further information.
We further looked for \textit{TESS} Input Catalog (TIC) v8.2 and CTL v8.01 catalog~\citep{TICv7,TICv8,TICv82}, 
and adopted the basic astronomical parameters in Table~\ref{tab:TESS-CTLv801}. 
%
%
\begin{table}[!thb]
  \caption[]{\textit{TESS} observations (at two-minute cadence) of HD 52788, HD 53166 and HD 53349. }
  \label{tab:TESS-data}
  \begin{center}\begin{tabular}{lcccll}
  \hline
\multicolumn{3}{l}{ Star Identifiers 	}                     & $V$      &Sp.  Type       \\
\multicolumn{3}{r}{V =HD 52788 = TIC 279361762} & 8\fm37 & Fm\_dD\\
\multicolumn{3}{r}{C1=HD 53166 = TIC 279431011} & 8\fm1  & A1V   \\
\multicolumn{3}{r}{C2=HD 53349 = TIC 279476396} & 6\fm0  & A8III or F0V\\
 \hline
 \multicolumn{6}{l}{Orbits;  Sector   ~~~~~~~~~~~~~~ 11--25, 61--85;2--10,12,13,27--39}\\
(Camera,CCD)	 &\multicolumn{5}{l}{\hspace{-14.95mm}(4,4),(4,1),(4,3),(4,2)}\\
&\multicolumn{5}{l}{\hspace{-15.95mm} S0036 started on 2021 Mar 1} \\
Time span &\multicolumn{5}{l}{ \hspace{-14.95mm}BJD 2458354.107--2459389.716} \\
\multicolumn{6}{r}{(2018 Aug 22 23:50--2021 May 26 02:43UT,1035.6 days)}\\
Data length &\multicolumn{5}{l}{ \hspace{-15.95mm} 24 sectors * 13.7 days * 2}\\
&\multicolumn{5}{l}{ = 657.599 days }\\
Frequency resolution &\multicolumn{5}{l}{ 0.0009656 \,d$^{-1}$ }\\
Number of data points analyzed&\multicolumn{5}{l}{ 419,915 }\\
\multicolumn{6}{r}{ \hspace{-10.95mm}(outliers and unusual data points excluded)}\\
Total number of data &\multicolumn{5}{l}{ 422,046}\\
\hline
  \end{tabular}\end{center}
\end{table}

\subsection{Custom Data Processing}

In the current work, we first did graphic screenings on each sector's light curves for choosing SAP or PDCSAP flux. 
A close visual inspection of both light curves plotted up and down and overlapping helped the author compare 
which one is better for pulsation detection.  
If the SAP flux looks no evident difference with PDCSAP flux, then use SAP flux.
If the PDCSAP flux data lost too many points while the SAP flux data are flat, then we choose SAP flux, 
otherwise, 
especially in the case of severe systematic trends showing slope and steep brightness, PDCSAP flux is chosen. 
Figure~\ref{fig:SAP-PDCSAP} shows two examples.  
In any SAP or PDCSAP flux, those data points flagged as ``bad quality" were removed. 

During a course of surveying a large amount of \textit{TESS} light curves the author found that both PDCSAP and SAP fluxes could be 
sometimes suffered from a systematic shift in brightness between the two orbits of a \textit{TESS} sector, 
which would result in abnormal discontinuity and artificial displacement of the two orbits' light curves. 
This is quite frequently occurred in HLSP-QLP (uncorrected SAP) light curves for long-period variables, for instance, Mira variables TIC 329891910, TIC 457021714, and TIC 359401246 on sector 38. 
Displacements of a segment light curve also occurred within individual orbits, e.g. RRLyr star TIC 127088233 on sector 38 and so on. 
Thus, a custom processing of correction to the displacement is applied whenever needed.
With the above criteria, these final used fluxes of PDCSAP and SAP mixture are ensured the \textit{TESS} best estimate of 
the intrinsic variability of a target.
Any remained low-frequency instrumental systematics would be either not removed or introduced by the \textit{TESS} reduction pipeline~\citep{cunha2019}. 

\begin{table}[!htb]
  \caption[]{Astronomical Parameters of HD 52788, HD 53166 and HD 53349. }
  \label{tab:TESS-CTLv801}
\setlength{\tabcolsep}{2.55pt}              
\renewcommand{\arraystretch}{1.005} 
  \begin{center}\begin{tabular}{llll}
  \hline
  Para.    &V=HD 52788 & C1=HD 53166  & C2=HD 53349 \\
 \hline
 $B$ & 8.788$\pm$0.028 &8.27$\pm$0.027  & 6.283$\pm$0.023   \\
 $V$ & 8.39$\pm$0.03   &8.15$\pm$0.03   & 6.01$\pm$0.03  \\ 
 $J$ & 7.683$\pm$0.024 &8.07$\pm$0.026  & 5.422$\pm$0.018  \\
 $H$ & 7.521$\pm$0.027 &8.045$\pm$0.024 & 5.296$\pm$0.023  \\
 $K$ & 7.473$\pm$0.029 &8.006$\pm$0.023 & 5.245$\pm$0.016  \\
 Tmag & 8.0396       &8.17            & 5.722\\
 Sp. type  & Fm\_dD       & A1V            & A8III \\
 $T_{\rm eff}$  & 6838$\pm$134.87  &9562$\pm$186.18  &  7118.48$\pm$111.54\\
 $\log g$          & 3.319$\pm$0.092   & 4.21$\pm$0.06      &  4.039$\pm$0.079   \\
 $R/R_{\odot}$ & 4.412$\pm$0.216& 2.026$\pm$0.053 &  1.988$\pm$0.069 \\
 $M/M_{\odot}$ & 1.48$\pm$0.25        & 2.43$\pm$0.32     & 1.59$\pm$0.25  \\
 $L/L_{\odot}$ &38.35  & 30.93 & 9.15 \\
 $M_V$          &0.987   & 1.304 & 2.335\\
 D (pc)	      &302.42 &234.1 & 54.3\\
\hline
  \end{tabular}\end{center}
\end{table}

Last, we proceed to process the \textit{TESS} data. 
The fluxes (SAP\_Flux and PDCSAP\_Flux)
corresponding magnitudes  (SAP\_mag and PDCSAP\_mag) are calculated using the 
formula  
$mag = -2.5\log(flux) + 20.4436 $. 
The magnitude zero point 
was selected for a resulting magnitude matching with  
Tmag of TIC v8.2. 
By arbitrarily selecting a star, TIC 41196013 (Tmag=11\fm4745) which was observed in \textit{TESS} sector 41 and confirmed as a non-variable
in a time scale of days for calibrating the magnitude zero point. 
We found 20.4436 and 20.2531\,mag for converting SAP and PDCSAP fluxes into magnitudes, respectively.
 In case of two successive orbits observations (each $\sim$13.7 days) 
having different mean values,  
a unique zero mean was applied for Fourier analysis.

Such flattening 
to a uniform level by subtracting individual means of each part was 
also applied to the uneven cases from sector to sector. 
Therefore light curves between two orbits and among sectors are always lined up before doing Fourier analysis.
This adjustment of magnitude 
zero points ensures the elimination of externally-entailed additional noise in the low-frequency domain.
 Concerning no atmospheric effect on the \textit{TESS} photometry, it is not necessary to establish differential magnitude as 
in traditional ground photometry. Thus the above-calculated magnitudes were used in our analyses. 
This way also allows us to diagnose each star without effects brought in by comparison stars.

\section{Frequency analysis}
We analyze pulsation contents by resolving light curves into a sum of multiple sine or cosine waveforms with 
different frequencies, amplitudes, and phases through classical Fourier transform. 
Fourier analysis was carried out by using {\sc period04}~\citep{lenz05}. 
It is known that noise in low-frequency domain, aliasing, and false peaks caused externally are three key issues which should be 
noticed in Fourier analysis. 
Caution was taken in picking up a frequency when strong low-frequency instrumental noise is present 
even though our custom processing of \textit{TESS} SAP and PDCSAP light curves data has exerted in minimization of external sources. 

It is well known that stellar activities may not be periodic, and even irregular.  
Periodic signals including pulsation are not strictly sinusoidal. Deviations of multiple sines to light curves need careful evaluation. Therefore, when the residuals of least-squares fitting of multiple frequencies to light curves improve insignificantly, we have to stop prewhitening.  Moreover, we pick up frequencies over a signal-to-noise (SNR) level of 4.3 following~\citet{1993A&A...271..482B}.  
We recalculated residuals and SNR with the remained significant frequencies.

Aliasing in Fourier spectrum is resulted from discrete data sampling. 
Ground astronomy has various sampling intervals and daytime gaps without data.
Uninterrupted space observations never mean no interruption in observations, they are still true discrete sampling in time-space.
\textit{TESS} exposed every two seconds, HDU stacked images into groups of 2-minute cadence and 30-minute cadence for different science goals. 
Data-downloading periods together with
orbits and sectors switch, these are actual \textit{TESS} data sampling intervals. 
Aliasing, as a data acquisition effect caused by sampling intervals is not yet completely eliminated in space-based data.
 
Here we explored the \textit{TESS} data aliasing. 
First, similar to~\citet{2019AJ....157..178S}, the data points 
showing severe residual uncorrected systematics in the fits are removed in the joint analysis presented in this paper.
We know that 
single-site ground-based photometry was affected by daily aliases, which occur when observations are not made during the day and the next night's observations begin almost 24 hours later.
Under Nyquist sampling theorem, frequencies less than half the sampling rate will not alias: 
$f_{\rm max}=1/(2\Delta t)$.
\textit{TESS} observations' sampling intervals are 2 and 30 minutes, 
that means the maximum possible frequencies (i.e. the Nyquist frequency) should be 
less than 360 and 24\,d$^{-1}$ , respectively.
However, besides the regular sampling intervals, there are additional time gaps 
in \textit{TESS} data sets between two consecutive orbits in a sector as well as between two successive sectors. 
The gaps in the middle of each sector's time series are due to the data downlink separating the two physical orbits within each \textit{TESS} sector.

\begin{figure}[!htb]
   \vspace{2mm}\centering
   \includegraphics[width=0.55\textwidth, height=45mm,angle=0,scale=0.95]{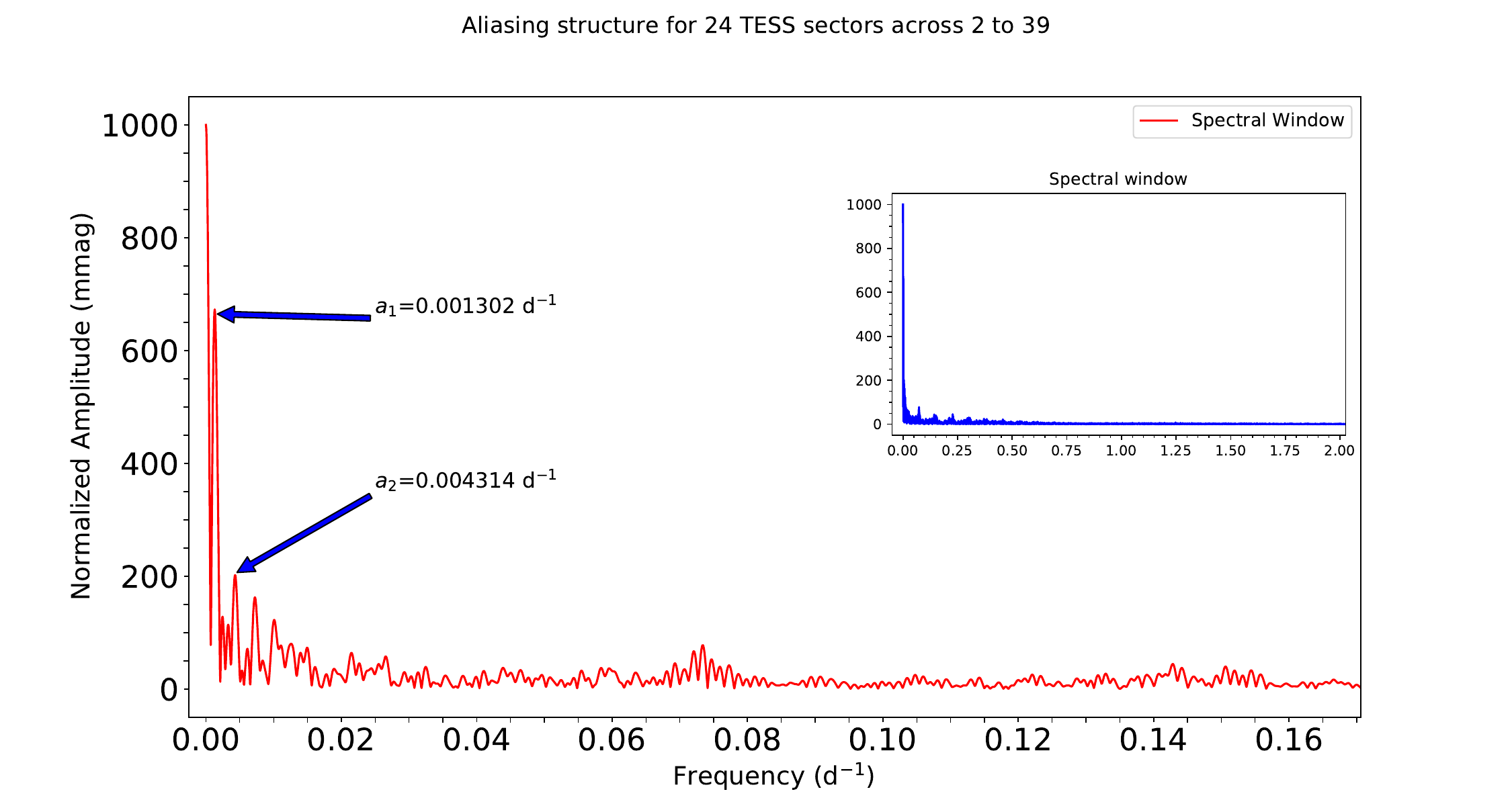} 
   \caption{Aliasing structure for the combination data of 24 \textit{TESS} sectors 2--10,12,13,27--39.}
   \label{fig:window-HD52788} 
\end{figure}

The current available \textit{TESS} data for the three targets cover 24 sectors across the 2nd to 39th sectors (see Table~\ref{tab:TESS-data}).
This set of data spanned 
1035.61 days, starting on BJD 2458354.10718 to 2459389.71673 (i.e. between 2018.08.23 14:33 and 2021.06.24 05:12).
Besides regular orbits and sectors switch intervals -- actually they are not fixed and they depend on the pipeline reduction of PDCSAP (roughly in 1.0--4.0 days in current case\footnote{TESS observing dates refer to: https://heasarc.gsfc.nasa.gov/docs/tess/sector.html}), there are no data available in sector 11 and sectors 14 through 26 over the entire 2nd to 39th sector period.  
As a composite effect, the spectral window reflects the real aliasing structure. 
Figure~\ref{fig:window-HD52788} shows the spectral window, which is much sharp.
Two stronger aliases are marginal in the zoomed graph at $a_1$=0.001302 and $a_2$=0.004314\,d$^{-1}$. 
Nyquist sampling theorem indicates that a frequency peak will be accompanied by the combination side peaks 
at $f\pm a_{1}$ and $f\pm a_2$. 
Due to the aliases are actually weak, each side peak will less contaminate intrinsic frequencies 
and will be not further aliased to harmonic spaces of either $a_1$ or $a_2$ in our case. 
Under the above scheme, the final derived periods have ejected all possible aliases. 

In addition, 
the frequency resolution of the time-series data is an important consideration when identifying closely spaced frequencies in periodograms. 
The frequency resolution of a Fourier transform denotes the smallest frequency difference that can be distinguished. 
A higher frequency resolution allows for better discrimination between closely spaced frequencies.
The frequency resolution is determined by the length of the time series and the sampling rate by the following formula:
$\Delta f = f_s / N $
where $f_s$ is the sampling rate and $N$ is the number of data points used in the FFT.
It follows that better frequency resolution can be achieved using a longer data sequence.
Assuming 2-minute cadences sampling rate with total 419915 data points analyzed, the resulting frequency resolution is 
$\Delta$f=0.001715\,d$^{-1}$. 

Alternatively, the frequency resolution is restricted by the total time base length ($T$) of a dataset following the traditional Rayleigh criterion and is theoretically defined by
$\Delta f = 1/T$, which 
would result $\Delta f$= 0.0009656\,d$^{-1}$ if we used the time span $T$=1035.61 days across over sectors 2 to 39.  
In current case, the effective time coverage with data 
is 24 sectors having 48 orbits, the actual time duration equals to $T_r\sim$13.7*48 = 657.6 days, 
that is 63.49\% of the above observing time span (refers to Table~\ref{tab:TESS-data}).  
Thus an alternative effective frequency resolution would take the value of $1/657.6$ = 0.00152068\,d$^{-1}$. 
This value implies a much conservative frequency resolution of about $1.6/T$, 
a coincidence with the argument of $1.5/T$ as the resolvable frequency spacing~\citep{1978Ap&SS..56..285L} . 
However, we learned that ~\citet{2008A&A...481..571K, 2017A&A...603A..13K} suggest a less conservative criterion of $1/(4T)$ based on Monte-Carlo simulation. 

Taking a further comparison with the \textit{Kepler} light curve of the $\delta$ Sct star KIC 7106205 which spans a total of 1470\,d (4\,yr) consisting of 65\,308 data points where the frequency resolution was $1/T$=0.00068\,d$^{-1}$~\citep{2014MNRAS.444.1909B}, 
if one used sample frequency divided by the number of data points as frequency resolution (i.e. $\Delta f=f_s/N$), 
then it would be only 0.0226\,d$^{-1}$ assuming the short cadence data. 
The later value is 33 times of the previous one.
In this regard, the above calculated value of $\Delta f = f_s / N$ =0.001715\,d$^{-1}$ (roughly equals to $2/T$) for the present 24 \textit{TESS} sectors data was over-estimated.  
It is a known shortcoming that the \textit{Kepler} short and long cadence data have integration times of 58.5\,s and 29.5\,min, respectively, not good for detection of short time-scale variations and close frequencies given limited frequency resolution~\citep{2011MNRAS.414.2413S, 2012MNRAS.422..665M}.
\textit{TESS} has better data coverage within a sector with shorter cadence sampling, 
the frequency resolution of 24 \textit{TESS} sectors is excellent and it's value should not be over-estimated.  

Now with these estimations used in the literature, we can reach a good compromise with a particular emphasis on continuous space photometric data. 
As the geometric mean is less affected by extreme values than the arithmetic mean, 
we propose to derive the weighted geometric mean of the three estimations $1/(4T)$, $1.5/T$ and $1/T$ with respective weights of 2, 0.5 and 1 calculated by
 \begin{equation}
 \Delta f_1 = \sqrt[^{3.5}\!]{(\frac{1}{4T})^2 * (\frac{1.5}{T})^{0.5} * \frac{1}{T}} \approx \frac{1}{2.1T} 
\end{equation}
and then the weighted geometric mean of the two extreme estimations $1/(4T)$ and $1.5/T$ with respective weights of 2 and 1 calculated by
\begin{equation}
\Delta f_2 = \sqrt[^3\!]{(\frac{1}{4T})^2 * \frac{1.5}{T}} \approx \frac{1}{2.2T} 
\end{equation}
Last take the arithmetic mean of the two geometric means, we get
\begin{equation} 
\Delta f = \overline{\Delta f_1 + \Delta f_2}  \approx \frac{1}{2.15T} 
\end{equation}

Assigning double weight  to $1/(4T)$ is a deliberate strategy in regard to the high accuracy and perfect time coverage of the 24 \textit{TESS} sectors data.  
So taking account of large amount of accurate well-sampled space data we are able to confidently adopt  a realistic estimation of  $\Delta f=1/(2.15T)$=0.000449\,d$^{-1}$
as the smallest distinguishable frequency difference when resolving two closely spaced pulsation contents for the 24 \textit{TESS} sectors data. 
Two close significant frequencies with a separation larger than 0.000449\,d$^{-1}$ will be taken as two individual 
frequencies, otherwise one frequency. 

Nevertheless, the frequency resolution criterion itself could dramatically over-estimate the real uncertainty in frequencies.
Frequency resolution is related to the error in frequencies by the formula
\begin{equation}
\sigma_f = \sqrt{\frac{6}{N}} \frac{1}{\pi T_r} \frac{\sigma(m)}{a}~~~
\end{equation}
where $a$ is amplitude, $N$ is the number of analyzed data points, $\sigma(m)$ is the root-mean-squares (rms) of a least-squares sinusoidal fit 
and $T_r\equiv N\Delta t$ is the total time length of the data ~\citep[see eq.(11) in][]{1999DSSN...13...28M}
other than the above observing time span in most cases.

For the Fourier parameters resulted from {\sc Period04}, which also 
calculated the analytical uncertainties on 
each frequency, amplitude, and phase of the best-fitting sinusoids following the formulae derived by 
both~\citet{1999DSSN...13...28M}
and~\citet{1999A&A...349..225B}.
We have assumed the residuals from the multiple frequencies least-squares fitting to be 
the root-mean-squares (rms) deviations of observational noise corresponding to the 24 sectors data set, 
i.e.  $\sigma(m)$=0.00273 mag. 
Learned from~\citet{2008A&A...481..571K} who explored the Fourier parameters errors in detail, 
the above determined errors would not be underestimated.

%

\section{Results}
We report in this section our results for each analyzed star in tables of frequency solution and in graphs of light curves, periodograms and prewhitened residuals. 

\subsection{HD 52788 = TIC 279361762}
\begin{figure}[!htb]
   \vspace{2mm}\centering
   \includegraphics[width=0.52\textwidth, height=92mm,angle=0,scale=0.99]{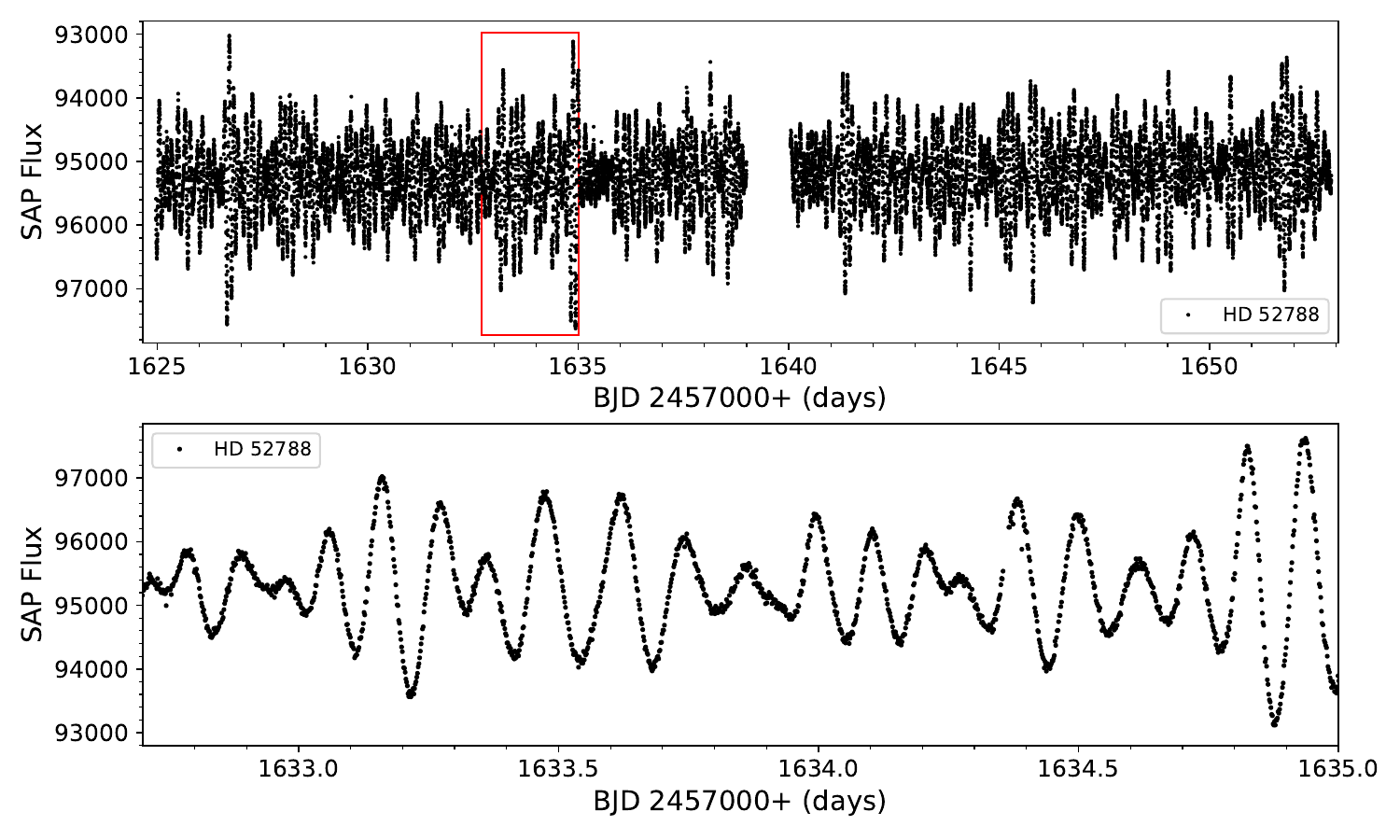} 
\par\vspace*{-3mm}  
\hspace*{-0mm}\includegraphics[width=0.52\textwidth, height=55mm,angle=0,scale=1.02]{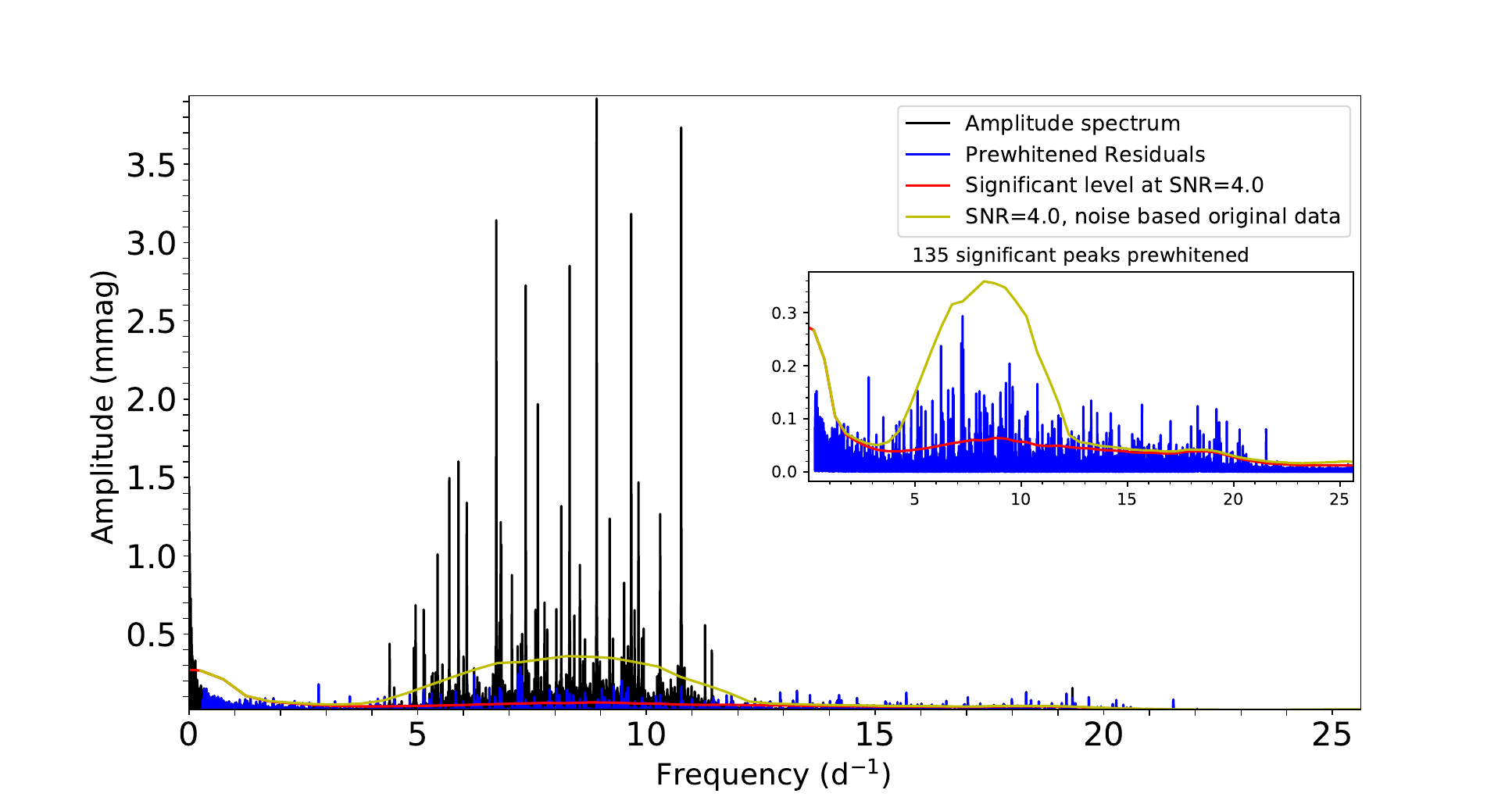}
   \caption[]{\textit{TESS} light curves of HD 52788=TIC 279361762 (Top: on sector 12; Middle: the red portion zoomed-in ) and the amplitude spectrum using all 24 sectors (2--10, 12,13, 27--39).  }
   \label{fig:52788-sp-SNR}
\end{figure}

\textit{TESS} observed HD 52788 in total of 24 sectors across 2 to 39 over 1035 days from 2018 August 22 through 2021 May 26.
Light curves are available as sources of \textit{TESS}-SPOC, HLSP-QLP, and HLSP-SPOC. 
The \textit{TESS}-SPOC 2-minute cadence SAP flux data were used.
We see the MIT HLSP-QLP team attempted resolving the data with multiple planets fittings, but no final prospective outcome.
The light curves exhibit a much more complex structure visually.
With successive prewhitenings and multiple frequencies fittings, the residuals are improved slightly (see Table~\ref{tab:52788-sigma}), more frequencies do not satisfactorily account for the complexity of the light curves of HD 52788.  
With a big number of parameters when simultaneous optimization for frequencies, amplitudes, and phases, 
non-linear least-squares fitting encountered a divergent issue of 'matrix cannot be inverted', which failed to improve the three parameters of the fitting. 
So we have to stop prewhitening procedure when reaching the 191st trial frequency.
Then we computed noise levels based on both residuals and original data for comparison of significant peaks.
Though up to 191 trial frequencies were prewhitened in the procedure,  
135 frequencies with significant signal-to-noise ratios over 4.3 were screened out. 
Then multiple frequencies least-squares fitting using the 135 frequencies was done, 
and the residuals, noise level and SNR were recalculated. 
We applied the frequency resolution adopted above of 0.000449\cd~to discriminate close frequencies 
and finally a total 126 individual  significant pulsation frequencies are remained, 
along with 3 combination terms and 9 significant frequencies under resolving limit. 
Frequency analysis shows that the residual spectrum even with 135 frequencies pre-whitened is still full of peaks 
in the range of 5--15\,d$^{-1}$ over the calculated significant curve using residuals but below that calculated from original data (see the insert of Fig.~\ref{fig:52788-sp-SNR}). 
Table~\ref{tab:52788-freq} reports the final results, 
where the errors for frequency and phase are given in a compact uncertainty notation way.
The number in parentheses represents the uncertainty of the measurement and it is  
expressed as one standard deviation ($1\,\sigma$).
Figure~\ref{fig:52788-sp-SNR} displays the light curves on a sector and periodogram based on all 24 sectors data.

\begin{table}[!htb]
  \caption[]{Residuals of Multi-frequency Least-Squares Fitting with Successive Prewhitenings.  }
  \label{tab:52788-sigma}
\setlength{\tabcolsep}{2.55pt}              
\renewcommand{\arraystretch}{1.005} 
  \begin{center}\begin{tabular}{ccc}
  \hline
Number of  & Residuals of & Improvement \\
Frequencies &  Fitting        & Rate\\
 \hline
\multicolumn{1}{c}{12}   & 0.0042389 & --\\
\multicolumn{1}{c}{32}   & 0.0032422 & 23.5\% \\
\multicolumn{1}{c}{79}   & 0.0030626 & 5.54\% \\
\multicolumn{1}{c}{108} & 0.0030232 & 1.29\% \\
\multicolumn{1}{c}{120} & 0.0029725 & 1.68\% \\
\multicolumn{1}{c}{158} & 0.0028472 & 4.22\% \\
\multicolumn{1}{c}{169} & 0.0028147 & 1.14\% \\
\multicolumn{1}{c}{191} & 0.0027301 & 3.01\% \\
 \hline
  \end{tabular}\end{center}
\end{table}

\begin{table}[!htb]
  \caption[]{Frequency solution of HD 52788 (=TIC 279361762) based on all available \textit{TESS} data secured in 24 Sectors.
The digits in parentheses represent the error in the last one or two decimal places. Amplitude in mmag and errors in amplitudes are 5.96 ppm.}
  \label{tab:52788-freq}
  \begin{center}
\setlength{\tabcolsep}{2.95pt} 
\renewcommand{\arraystretch}{1.005} 
  \begin{tabular}{lclr}
\hline\noalign{\smallskip}
Frequency (d$^{-1}$)   &     Amplitude  & Phase (0-1) & SNR \\ 
 \hline\noalign{\smallskip}
$f_{0}$=10.760712(1)  &  3.757   &  0.6532(3) & 299.1 \\
$f_{1}$= 8.912941(1)  &   3.336   &  0.4876(3) & 208.7 \\
$f_{2}$= 6.718369(1)  &   3.145   &  0.5542(4) & 233.8 \\
$f_{3}$= 8.322361(1)  &   2.863   &  0.0899(3) & 193.2 \\
$f_{4}$= 7.361251(2)  &   2.730   &  0.5630(6) & 191.2 \\
$f_{5}$= 9.667860(3)  &   2.711   &  0.6036(9) & 188.3 \\
$f_{6}$= 7.625737(2)  &   1.647   &  0.4090(6) & 109.2 \\
$f_{7}$= 9.827841(2)  &   1.635   &  0.0752(7) & 113.6 \\
$f_{8}$= 5.891104(2)  &   1.634   &  0.9169(6) & 142.2 \\
$f_{9}$= 9.940035(3)  &   1.520   &  0.5681(8) & 105.6 \\
$f_{10}$= 5.692326(3)  &   1.507   &  0.3937(9) & 131.2 \\
$f_{11}$= 8.912464(3)  &   1.483   &  0.5398(7) & 92.7 \\ 
$f_{12}$= 8.139382(2)  &   1.330   &  0.0989(7) & 89.8 \\
$f_{13}$= 9.197316(2)  &   1.324   &  0.7725(7) & 84.2 \\
$f_{14}$= 6.072851(2)  &   1.311   &  0.0325(6) & 105.1 \\
$f_{15}$= 6.816985(3)  &   1.264   &  0.1831(7) & 94.0 \\
$f_{16}$=10.301894(3) &   1.257   &  0.0196(8) & 89.7 \\
$f_{17}$= 6.828059(4)  &   1.200   &  0.4087(11) & 89.2 \\
$f_{20}$= 5.434878(3)  &   1.012   &  0.1317(10) & 94.3 \\
$f_{21}$= 8.546385(2)  &   0.980   &  0.9524(6) & 61.3 \\
$f_{23}$= 7.260373(3)  &   0.968   &  0.7540(9) & 67.8 \\
$f_{25}$= 7.059142(5)  &   0.897   &  0.8382(16) & 62.8 \\
$f_{27}$= 9.512210(7)  &   0.799   &  0.5763(22) & 55.5 \\
$f_{28}$= 7.772967(15)  &   0.754   &  0.6716(44) & 50.0 \\
$f_{30}$= 7.575876(4)  &   0.693   &  0.4292(12) & 45.9 \\
$f_{31}$= 4.954131(6)  &   0.687   &  0.4062(19) & 67.8 \\
$f_{32}$= 8.032471(4)  &   0.686   &  0.7784(11) & 46.3 \\
$f_{33}$= 9.598536(7)  &   0.678   &  0.6018(21) & 47.1 \\
$f_{34}$= 8.425755(4)  &   0.609   &  0.2449(12) & 41.1 \\
$f_{35}$= 11.284097(6) &  0.544   &  0.4131(17) & 44.9 \\
$f_{36}$= 6.786902(6)  &   0.540   &  0.3683(18) & 40.1 \\
$f_{37}$= 7.834032(5)  &   0.494   &  0.2679(14) & 32.7 \\
$f_{38}$= 9.269924(7)  &   0.479   &  0.9459(20) & 30.4 \\
$f_{39}$= 9.515054(7)  &   0.466   &  0.2663(20) & 32.4 \\
$f_{41}$= 9.743959(11) &   0.430   &  0.7088(315) & 29.8 \\
$f_{42}$= 4.386572(7)  &   0.428   &  0.0939(22) & 44.2 \\
$f_{43}$= 9.191466(8)  &   0.424   &  0.2348(22) & 27.0 \\
$f_{44}$= 5.130545(5)  &   0.409   &  0.4008(14) & 38.1 \\
$f_{45}$= 5.133285(8)  &   0.404   &  0.3003(23) & 37.6 \\
$f_{46}$= 10.772772(8) &   0.396   &  0.6975(24) & 31.5 \\
$f_{47}$= 4.919064(8)  &   0.395   &  0.2621(23) & 38.9 \\
$f_{48}$= 11.428913(8) &   0.392   &  0.3759(24) & 32.4 \\
$f_{49}$= 9.510808(8)  &   0.381   &  0.1872(25) & 26.4 \\
$f_{50}$= 5.156588(9)  &   0.368   &  0.5539(26) & 34.3 \\
  \noalign{\smallskip}\hline 
  \end{tabular}
  \end{center}
\end{table}

\addtocounter{table}{-1}
\begin{table}
\vspace{-5mm}
\caption{~~ --- Continued}\smallskip
\centering
\setlength{\tabcolsep}{2.55pt}              
\renewcommand{\arraystretch}{1.005} 
  \begin{tabular}{lccr}
\hline\noalign{\smallskip}
Frequency (d$^{-1}$)   &     Amplitude   & Phase (0-1) & SNR \\ 
 \hline\noalign{\smallskip}
$f_{51}$= 6.001676(9)  &   0.364   &  0.9169(26) & 29.2 \\
$f_{52}$= 7.256143(9)  &   0.361   &  0.2173(26) & 25.3 \\
$f_{53}$= 5.540744(9)  &   0.340   &  0.9782(28) & 29.6 \\
$f_{54}$= 8.657510(10)  &   0.339   &  0.3604(28) & 21.2 \\
$f_{55}$= 8.613910(10)  &   0.313   &  0.2644(31) & 19.6 \\
$f_{56}$= 10.271927(10) &   0.309   &  0.7930(31) & 22.0 \\
$f_{57}$= 9.458759(11)  &   0.303   &  0.4938(31) & 19.3 \\
$f_{58}$= 7.245365(11)  &   0.303   &  0.1817(32) & 21.2 \\
$f_{59}$= 11.281291(10)  &   0.302   &  0.6374(31) & 25.0 \\
$f_{60}$= 5.906064(11)  &   0.291   &  0.8418(32) & 25.3 \\
$f_{61}$= 7.184681(11)  &   0.287   &  0.1920(33) & 20.1 \\
$f_{62}$= 9.905449(11)  &   0.278   &  0.8832(34) & 19.3 \\
$f_{63}$= 11.285506(11)  &   0.278   &  0.7845(34) & 22.9 \\
$f_{64}$= 9.908260(12)  &   0.269   &  0.6558(35) & 18.7 \\
$f_{65}$= 6.233235(12)  &   0.265   &  0.2720(36) & 21.3 \\
$f_{66}$= 7.286575(12)  &   0.263   &  0.6753(36) & 18.4 \\
$f_{67}$= 5.393172(12)  &   0.255   &  0.9920(37) & 23.8 \\
$f_{68}$= 8.661686(13)  &   0.251   &  0.0729(38) & 15.7 \\
$f_{69}$= 8.663166(13)  &   0.242   &  0.3133(40) & 15.1 \\
$f_{70}$= 9.741146(13)  &   0.238   &  0.9258(40) & 16.6 \\
$f_{71}$= 5.330562(14)  &   0.235   &  0.7490(40) & 21.9 \\
$f_{72}$= 9.745367(14)  &   0.233   &  0.0881(41) & 16.2 \\
$f_{73}$= 8.350984(14)  &   0.226   &  0.4041(42) & 15.2 \\
$f_{74}$=11.431706(8)  &   0.216   &  0.1647(24) & 17.9 \\
$f_{75}$=11.427492(16)  &   0.201   &  0.0345(47) & 16.6 \\
$f_{76}$= 6.821376(16)  &   0.196   &  0.0796(49) & 14.6 \\
$f_{77}$=10.363481(17)  &   0.191   &  0.7322(49) & 13.6 \\
$f_{78}$= 8.101823(17)  &   0.187   &  0.2150(51) & 12.6 \\
$f_{79}$= 2.833634(18)  &   0.180   &  0.5280(52) & 15.0 \\
$f_{80}$= 9.552320(18)  &   0.175   &  0.3082(55) & 12.1 \\
$f_{81}$= 5.304389(19)  &   0.167   &  0.4670(56) & 15.5 \\
$f_{82}$= 7.393222(19)  &   0.167   &  0.1204(56) & 11.7 \\
$f_{83}$= 6.911048(20)  &   0.159   &  0.1184(59) & 11.8 \\
$f_{84}$=10.233948(22)  &   0.145   &  0.1224(65) & 10.4 \\
$f_{85}$=10.236667(22)  &   0.145   &  0.0489(66) & 10.3 \\
$f_{86}$= 6.573550(22)  &   0.143   &  0.8390(65) & 10.6 \\
$f_{87}$= 8.035146(23)  &   0.139   &  0.9151(68) & 9.4 \\
$f_{88}$= 6.305248(23)  &   0.138   &  0.5002(68) & 11.1 \\
$f_{89}$= 9.517827(23)  &   0.136   &  0.6075(69) & 9.5 \\
$f_{90}$= 13.292368(24)  &   0.135   &  0.1017(70) & 12.2 \\
$f_{91}$= 8.654768(24)  &   0.130   &  0.9650(71) & 8.1 \\
$f_{92}$=18.304428(25)  &   0.126   &  0.3905(74) & 13.0 \\
$f_{93}$= 7.838867(25)  &   0.125   &  0.7913(75) & 8.3 \\
$f_{95}$= 8.211857(26)  &   0.124   &  0.1116(77) & 8.3 \\
$f_{96}$= 12.925388(26)  &   0.123   &  0.3218(78) & 11.0 \\
$f_{97}$= 8.107359(26)  &   0.122   &  0.4421(77) & 8.2 \\
$f_{98}$= 8.105974(26)  &   0.121   &  0.0219(76) & 8.2 \\
$f_{100}$=19.184598(27)  &   0.119   &  0.5267(81) & 13.3 \\
\hline 
  \end{tabular}
\end{table}

\addtocounter{table}{-1}
\begin{table}
\vspace{-5mm}
\caption{~~ --- Continued}\smallskip
\centering
\setlength{\tabcolsep}{2.65pt}              
\renewcommand{\arraystretch}{1.005} 
  \begin{tabular}{lccc}
\hline\noalign{\smallskip}
Frequency (d$^{-1}$)   &     Amplitude   & Phase (0-1) & SNR \\ 
 \hline\noalign{\smallskip}
$f_{101}$= 7.565613(27)  &   0.118   &  0.0656(80) & 7.8 \\
$f_{102}$=10.552950(27)  &   0.117   &  0.8282(81) & 9.3 \\
$f_{103}$= 4.487832(27)  &   0.117   &  0.8186(81) & 12.1 \\
$f_{104}$= 9.029008(27)  &   0.115   &  0.2003(81) & 7.3 \\
$f_{105}$= 13.577598(28)  &   0.113   &  0.5387(84) & 11.0 \\
$f_{106}$= 14.213484(28)  &   0.111   &  0.1584(85) & 11.0 \\
$f_{107}$= 19.313794(30)  &   0.106   &  0.6342(89) & 11.8 \\
$f_{108}$= 3.516773(31)  &   0.103   &  0.3656(93) & 10.6 \\
$f_{109}$= 6.346939(31)  &   0.102   &  0.0902(93) & 8.1 \\
$f_{110}$= 17.029084(34)  &   0.097   &  0.4436(100) & 11.0 \\
$f_{112}$= 9.902661(34)  &   0.094   &  0.608(10) & 6.6 \\
$f_{113}$= 7.280701(35)  &   0.092   &  0.170(11) & 6.4 \\
$f_{114}$= 5.124742(35)  &   0.091   &  0.246(11) & 8.5 \\
$f_{115}$= 7.190440(36)  &   0.090   &  0.944(11) & 6.3 \\
$f_{116}$= 10.403170(36)  &   0.087   &  0.412(11) & 6.2 \\
$f_{117}$= 14.603881(37)  &   0.087   &  0.262(11) & 8.9 \\
$f_{118}$= 17.990261(36)  &   0.086   &  0.872(11) & 9.1 \\
$f_{119}$= 7.931700(38)  &   0.082   &  0.469(12) & 5.4 \\
$f_{120}$= 8.320538(38)  &   0.082   &  0.353(11) & 5.5 \\
$f_{122}$= 20.274321(41)  &   0.081   &  0.286(12) & 13.0 \\
$f_{123}$= 9.461381(40)  &   0.080   &  0.145(12) & 5.1 \\
$f_{124}$= 18.411317(47)  &   0.076   &  0.992(14) & 7.9 \\
$f_{125}$= 9.452980(42)  &   0.075   &  0.797(13) & 4.8 \\
$f_{126}$= 10.784904(44)  &   0.074   &  0.361(13) & 5.9 \\
$f_{127}$= 7.258622(46)  &   0.070   &  0.045(14) & 4.9 \\
$f_{128}$= 18.580695(47)  &   0.069   &  0.630(14) & 7.2 \\
$f_{129}$= 7.363065(49)  &   0.069   &  0.933(15) & 4.8 \\
$f_{130}$= 9.602568(50)  &   0.063   &  0.661(15) & 4.4 \\
$f_{131}$= 10.771011(51)  &   0.062   &  0.099(15) & 5.0 \\
$f_{132}$= 20.195160(59)  &   0.058   &  0.965(17) & 9.4 \\
$f_{133}$= 6.575096(57)  &   0.058   &  0.866(17) & 4.3 \\
\hline 
\multicolumn{4}{c}{Frequencies under resolving limit and combination terms}\\  
\multicolumn{4}{l}{$f_{18}$= $f_5$+0.000077= 9.667937(3)  ~~~~~   1.152  ~~~  0.2479(10)~~ 80.0}\\
\multicolumn{4}{l}{$f_{19}$= $f_9-$0.000078= 9.939957(3)  ~~~~~   1.120   ~~~  0.2602(8) ~~~ 77.8} \\ 
\multicolumn{4}{l}{$f_{22}$= $f_0$+0.000410=10.761122(1) ~~~~   0.973   ~~~  0.1426(3) ~~~ 77.5 }\\ 
\multicolumn{4}{l}{$f_{24}$= $f_{23}$+0.000013=7.260386(3)  ~~~~   0.946  ~~~  0.2253(9) ~~~ 66.2} \\
\multicolumn{4}{l}{$f_{26}$= $f_6$+0.000273= 7.626010(5)  ~~~~~   0.894   ~~~  0.6858(14) ~ 59.3} \\ 
\multicolumn{4}{l}{$f_{29}$= $f_7-$0.000116= 9.827725(5)  ~~~~~   0.693   ~~~  0.6135(14) ~ 48.2} \\ 
\multicolumn{4}{l}{$f_{40}$= $f_{33}-$0.000042= 9.598494(7)  ~~~~~   0.448   ~~~  0.1664(21) ~ 31.1} \\ 
\multicolumn{4}{l}{$f_{99}$= $f_{25}-$0.000338= 7.058804(26)  ~~~   0.120   ~~~  0.7787(78) ~~ 8.4} \\ 
\multicolumn{4}{l}{$f_{134}$= $f_{51}$+0.000268= 6.001944(55)  ~~~   0.057   ~~~  0.665(16) ~~ 4.6} \\ 
\multicolumn{4}{l}{$f_{94}$= $f_3 + f_4 -$0.000079=15.683533   ~~~  0.125   ~~~  0.984184  ~~~ 13.8} \\
\multicolumn{4}{l}{$f_{111}$= $f_0 + f_1 -$0.000122=19.673531  ~~  0.095   ~~~  0.851158  ~~ 12.5} \\
\multicolumn{4}{l}{$f_{121}$= $2f_0 -$0.000011= 21.521412   ~~~~~~  0.081   ~~~  0.819591  ~~~ 21.1} \\
\noalign{\smallskip}\hline 
\multicolumn{4}{l}{Zeropoint:  7.81185449 mag} \\ 
\multicolumn{4}{l}{Residuals:  0.00273008 mag} \\ 
  \noalign{\smallskip}\hline 
  \end{tabular}
\end{table}

\subsection{HD 53166}
\begin{figure}[!htb]
\vspace*{-8mm}\centering
\par\vspace*{-0.9mm}\includegraphics[width=0.52\textwidth, height=42mm,angle=0,scale=1.04]{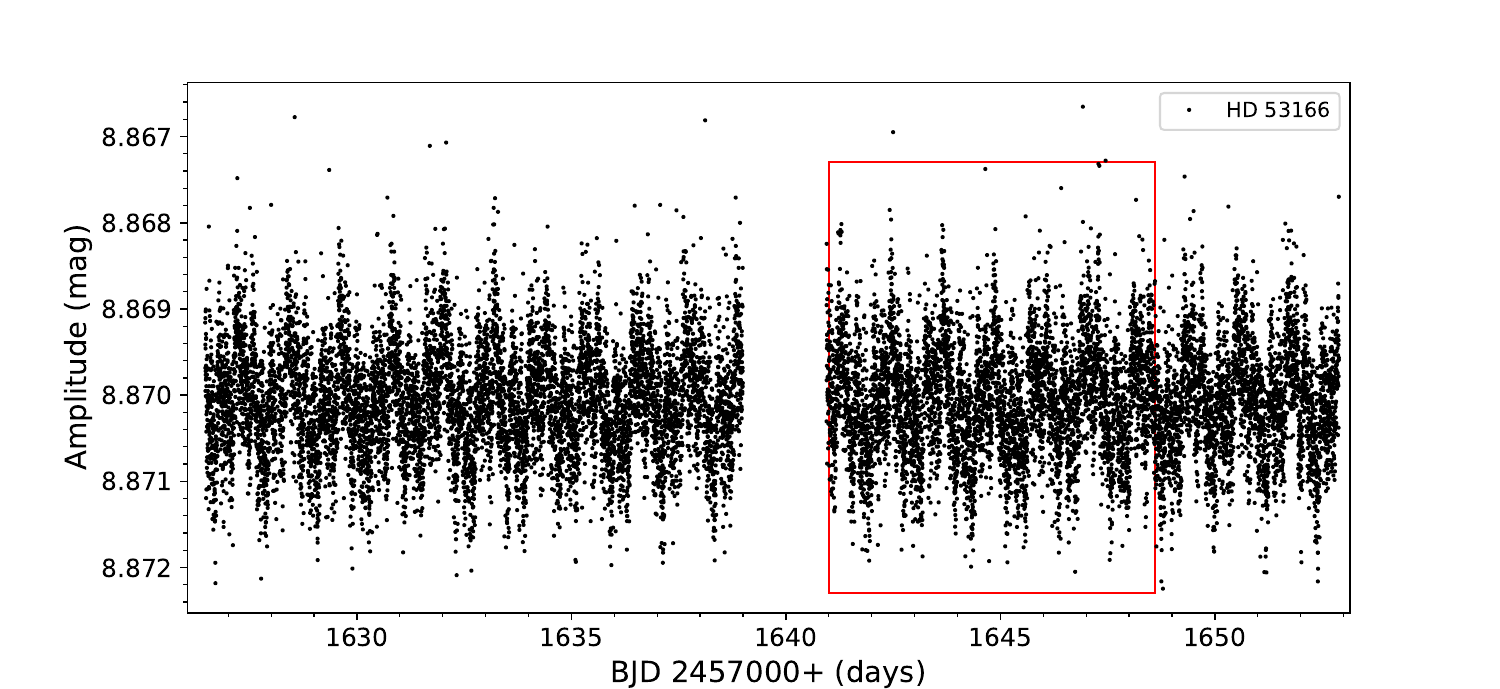}
\par\vspace*{-0.3mm}
   \includegraphics[width=0.52\textwidth, height=35mm,angle=0,scale=1.02]{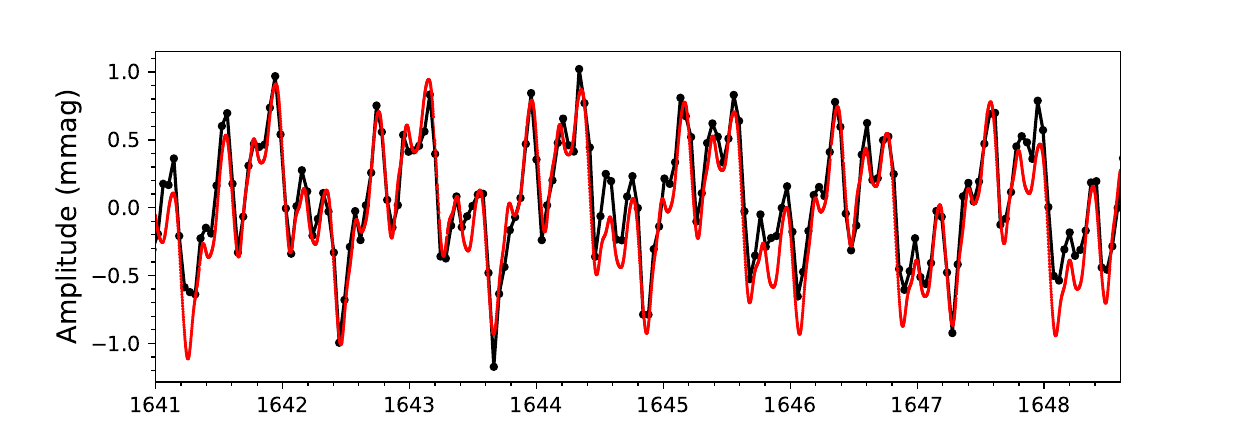}
\par\vspace*{-0.5mm}
\hspace*{-15pt}   \includegraphics[width=0.52\textwidth, height=50mm,angle=0,scale=1.00]{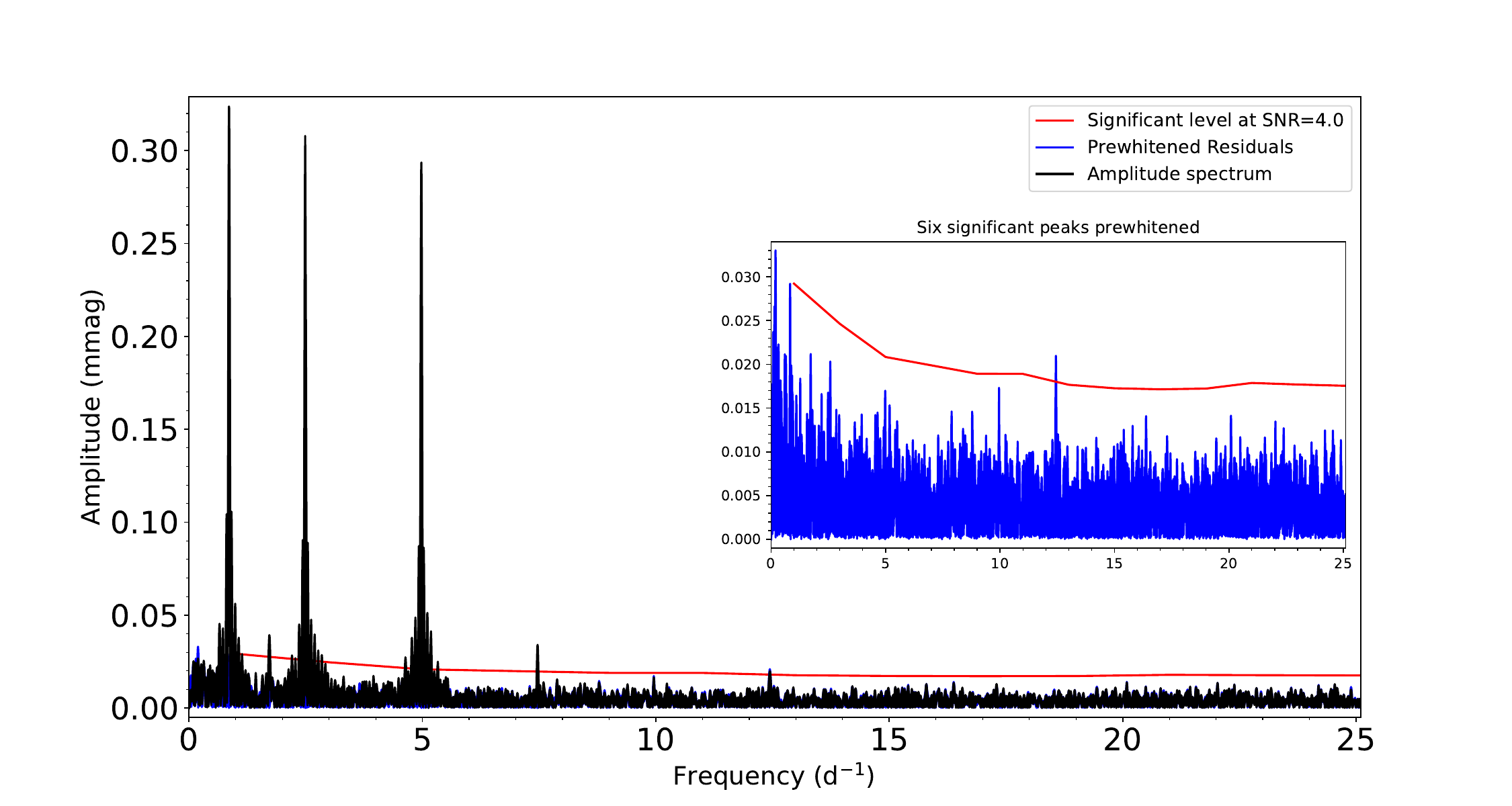}
\par\vspace*{-1.5mm}
\hspace*{-20pt}   \includegraphics[width=0.52\textwidth, height=50mm,angle=0,scale=1.03]{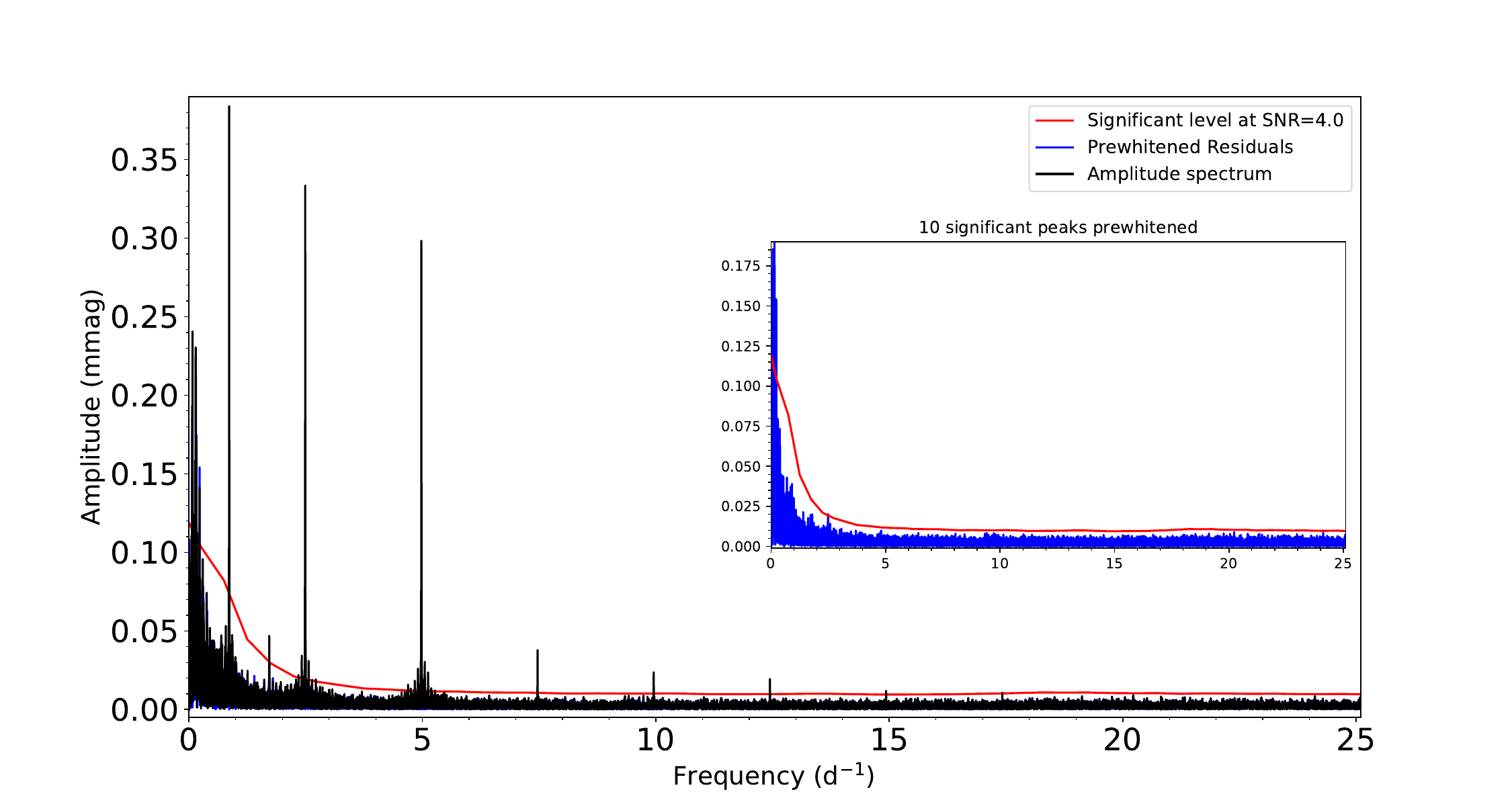}
   \caption[]{Light curves and amplitude spectrum of HD 53166=TIC 279431011. 
 Top: on sector 12; second panel: the red portion zoomed-in with fitting in red; third panel: 2-minute cadence on sectors 2--12; 
 bottom: 10-minute cadence on sectors 27--39.  }
   \label{fig:53166-sp-SNR}
\end{figure}

HD 53166 (=TIC 279431011, $V$=8\fm17, A1V, RA=07:00:34.60, Dec=$-$58:23:36.26 J2000.0), 
its \textit{TESS} light curves are available on sectors 2--13 in both 2-minute cadence (SPOC products) and 30-minute cadence (TASOC,TESS-SPOC, 
and QLP), and on sectors 27--39 as HLSP-QLP products in cadence of 10-minute. Total 24 sectors.
With \textit{TESS} data we reveal the star seems to be a new $\gamma$ Doradus and $\delta$ Scuti hybrid pulsating variable star.
Pulsation frequency contents are given in Table~\ref{tab:53166-freq}. 
Figure~\ref{fig:53166-sp-SNR} shows the light curves on sector 12 and 
periodograms for the 2-minute cadence data on sectors 2--12 together with the 10-minute cadence data on sectors 27--39, respectively.

\begin{table}[bht]
   \vspace{-40mm}
  \caption[]{Frequency solution of HD 53166 (=TIC 279431011) based on 24 \textit{TESS} sectors 2--13 and 27--39. 
A frequency term within
the frequency resolution 0.00152068\,d$^{-1}$ relative to one of the harmonics of 
the main frequencies or their combinations will be taken as real, otherwise an independent frequency. 
Digits in parentheses refer to errors in the last decimal place. Amplitude in mmag and errors in amplitudes are 2.02 ppm.}
  \label{tab:53166-freq}
  \begin{center}
\setlength{\tabcolsep}{1.95pt} 
\renewcommand{\arraystretch}{1.005} 
  \begin{tabular}{lccr}
  \hline\noalign{\smallskip}
Frequency (d$^{-1}$)   &     Amplitude  & Phase (0-1) & SNR \\ 
 \hline\noalign{\smallskip}
$f_{0}$= 0.85999(1)  &   0.326   &  0.265(1) & 19.8 \\
$f_{1}$= 2.48885(1)  &   0.307   &  0.804(1) & 44.27 \\
$f_{3}$= 0.85698(3)  &   0.120   &  0.250(3) & 7.30 \\
\hline
$f_{2}$= 2$f_1 -$  0.000126   &  0.301   &  0.297  & 78.67 \\
$f_{4}$= 2$f_0 +$  0.003006   &  0.049   &  0.921  & 5.84 \\
$f_{5}$= 3$f_1 -$  0.000203   &  0.028   &  0.752  & 11.32 \\
$f_{6}$= 4$f_1 -$  0.000344   &  0.023   &  0.139  & 10.41 \\
$f_{7}$= 5$f_1 -$ 0.000203  &  0.018   &  0.916 & 7.80 \\
$f_{8}$= 6$f_1 -$ 0.000595  &  0.016  & 0.509  & 6.84 \\
\hline 
\multicolumn{4}{l}{Zeropoint:  -4.43230403e-06 mag} \\ 
\multicolumn{4}{l}{Residuals:  0.000590436668 mag} \\ 
  \noalign{\smallskip}\hline 
  \end{tabular}
  \end{center}
\end{table}

\subsection{HD 53349 }
\begin{figure}[!htb]
   \vspace{2mm}\centering
\hspace*{-6mm}   \includegraphics[width=0.52\textwidth, height=55mm,angle=0,scale=1.09]{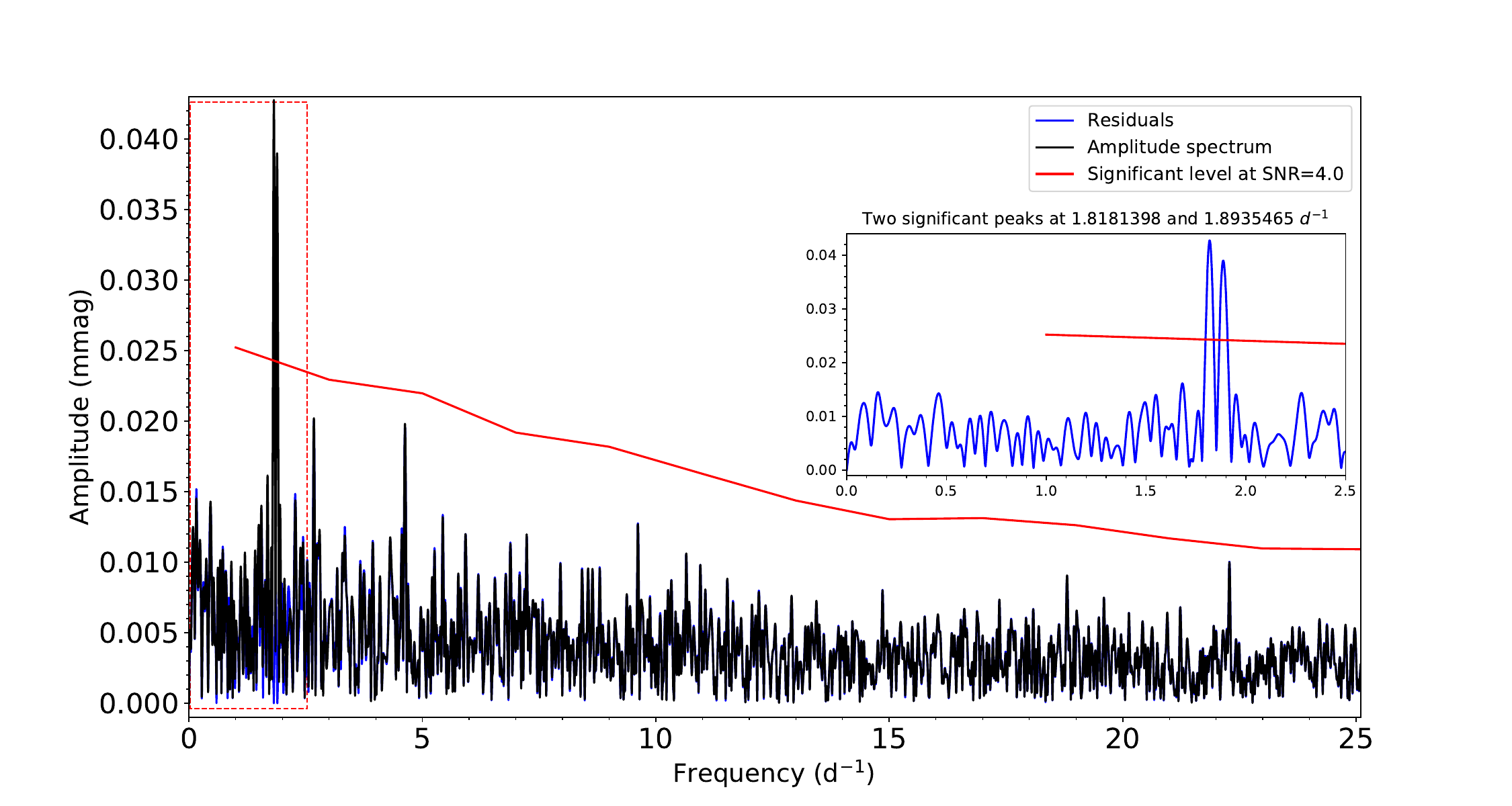}
\hspace*{-20pt}\includegraphics[width=0.52\textwidth, height=55mm,angle=0,scale=1.10]{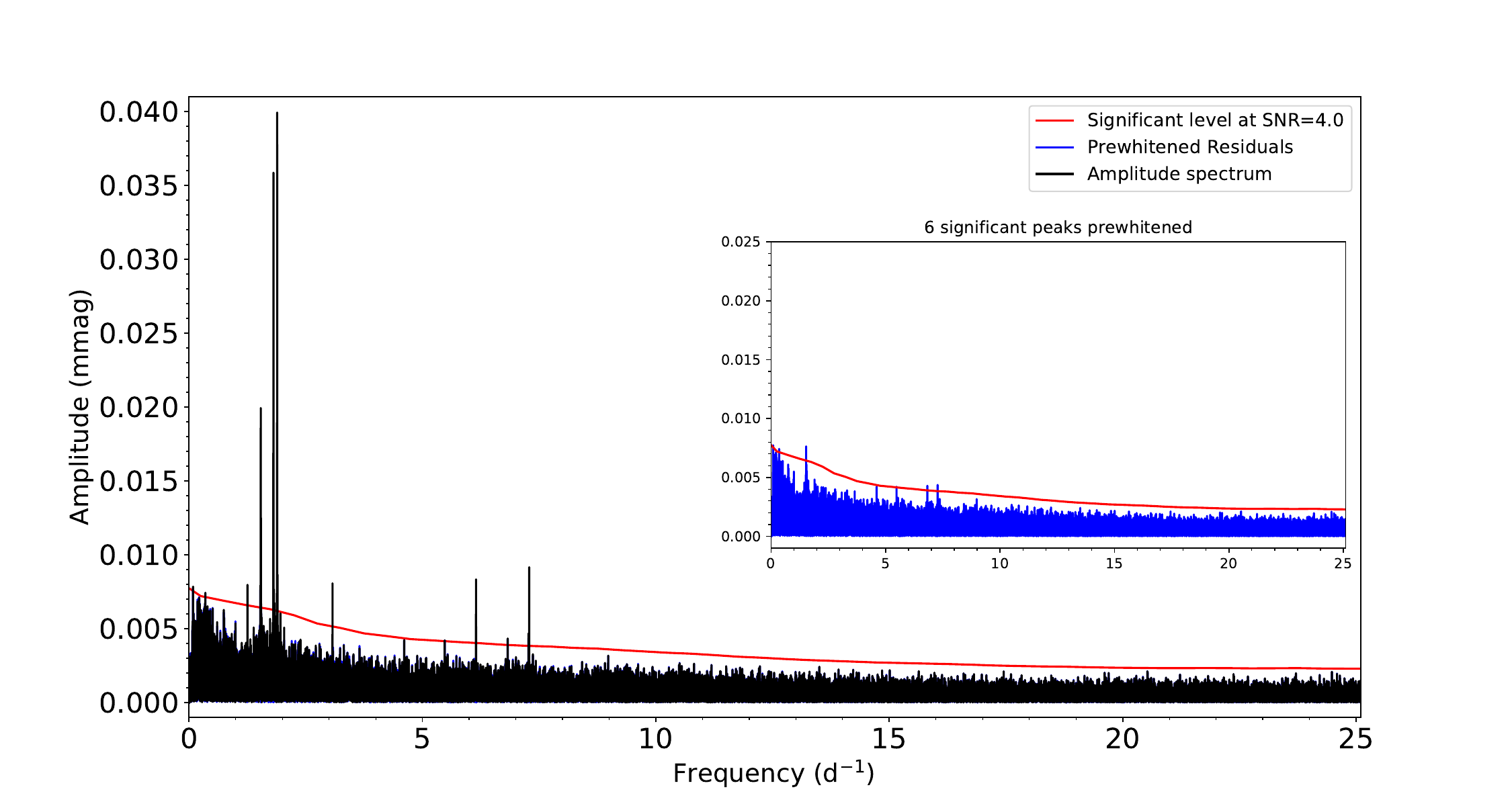}
   \caption[]{Amplitude spectrum of HD 53349 (=TIC 279476396). Top: sector 36, bottom: 24 sectors. 
}
   \label{fig:6396-sp}
\end{figure}

HD 53349 (=HR 2662 = TIC 279476396, $V$=6\fm0, A8III, RA=07:01:05.11 Dec=$-$58:56:23.77) is 
a high proper motion star. 
It was listed as NSV 3349 with F0V spectral type in the International Variable Star Index (VSX) of AAVSO but 
without an identified variability type\footnote{https://www.aavso.org/vsx/index.php?view=search.top}. 
Actually, the light variation is hardly visible from any single sector data. 
Analyses based on both the sector 36th and all available 24 sectors between 1st and 39th consistently lead to multi-periodic pulsational variations with two dominant frequencies at $f_1$=1.888139 and  
$f_2$=1.809503\,d$^{-1}$. 
Figure~\ref{fig:6396-sp} shows the 
periodograms for the 36th sector and the 24 sectors out of 1--39 at 2-minute cadence, respectively.
Frequency solution is reported in Table~\ref{tab:6396-freq}.
\begin{table}[!htb]
\begin{threeparttable}
   \vspace{-82mm}
  \caption[]{Frequency solution of HD 53349 (=TIC 279476396) Based on 24 \textit{TESS} sectors data.
Digits in parentheses refer to errors in the last two decimal places. Errors in amplitudes are 0.42 ppm.}
  \label{tab:6396-freq}
  \begin{center}
\setlength{\tabcolsep}{2.95pt}              
\renewcommand{\arraystretch}{1.005} 
  \begin{tabular}{lccr}
  \toprule
Frequency (d$^{-1}$)   &     Amplitude (mmag)  & Phase (0-1) & SNR \\ 
 \hline\noalign{\smallskip}
$f_{0}$= 1.888139(05)  &   0.039   &  0.4767(17) & 24.90 \\
$f_{1}$= 1.809503(06)  &   0.035   &  0.2293(19) & 22.35 \\
$f_{2}$= 1.538036(12)\tnote{*}  &   0.018   &  0.0002(36) & 11.58 \\
$f_{3}$= 7.287296(23)  &   0.009   &  0.7736(72) & 9.57 \\
$f_{4}$= 6.148190(26)  &   0.008   &  0.1311(79) & 8.34 \\
$f_{5}$= 3.074502(27)  &   0.008   &  0.4585(82) & 6.43 \\
\hline 
\multicolumn{4}{l}{Zeropoint:  4.43898411e-07 mag} \\ 
\multicolumn{4}{l}{Residuals:  0.000186609086 mag} \\ 
\bottomrule
  \end{tabular}
\begin{tablenotes}
     \small\footnotesize
      \item[*]   ~~~Accompanied with an equal amplitude peak at $f_{2} -$0.000794, 
      separated\\ \hspace*{4mm}over the adopted empirical frequency resolution of 0.000449\,d$^{-1}$, 
      but \\ \hspace*{4mm}within the theoretic frequency resolution of 0.0009656\,d$^{-1}$.
    \end{tablenotes}
  \end{center}
  \end{threeparttable}
\end{table}

\section{Discussion and Conclusion}
\label{summary}

With 24 \textit{TESS} sectors consisting of 48 orbits, each orbit $\sim$13.7-day uninterrupted high-precision photometry we are able to insight the real nature of the light variations of the known $\delta$ Sct star HD 52788 (=V383 Car). 
Surprisedly, we reveal that 
the two comparison stars HD 53166 and HD 53349 used in the only existing differential photometry of HD 52788 in the literature are now identified to be $\gamma$ Dor and $\delta$ Sct hybrid variable stars. 
We unveiled the variability types and pulsation contents for the three targets down to an unprecedented precision of 
pulsational amplitudes 5.7,  16, and 8 $\mu$mag, respectively.
 
\subsection{HD 52788}
Compared with previous ground results, the primary frequency 
$f_0$=10.765\,d$^{-1}$ is stable. The two old frequencies around 9.688 and 8.885 are changed and 
they should be blended mixture of multiple unresolved close frequencies. 
Surely a few days of the old dataset is too constrained to allow a good frequency resolving power 
as the present 24 sectors data spanning 1035.6 days with 
actual data length of 657.599 days which indicate a theoretic frequency resolution of 0.0009656\,d$^{-1}$ 
and an effective frequency resolution of 0.00152068\,d$^{-1}$, respectively. 
Instead of using these two much conservative thresholds -- $1/T$ and $1.6/T$, based on literature data, 
we derived an empirical frequency resolution of $\Delta f= 1/(2.15T)$=0.000449\cd as a discriminating criterion for 
distinguishing closely spaced frequency contents accounting for the large amount of accurate space data.	  
Because the pulsational contents of HD 52788 are mainly in the range of 4--12\,d$^{-1}$, 
and the three strongest frequencies of both HD  53166 and HD 53349 are lower than 5\,d$^{-1}$, 
HD 52788 should have not been significantly affected by the two comparison stars involved in 
the differential photometry in the previous studies. 
The amplitudes of C2=HD 53349 are so small (smaller than 60 micro-magnitude) that 
they would not influence in the results of the analysis of the light curves obtained with the old ground-based photometric observations of HD52788.
However, in the case of C1=HD 53166, the amplitudes are somewhat larger. 
HD 53166 could have had some influence on the primary peaks of HD52788 in the previous studies.
The pulsation frequency spectrum of HD 52788 over 126 individual frequencies is outstanding and rare among the $\delta$ Sct stars.  
Such a good number of pulsation frequencies deserve a follow-up asteroseismic analysis in a separate work.

In~\citet{2005A&A...435..955B}, 
75+ frequencies for the famous $\delta$ Sct star FG Virginis were resolved using extensive photometric multisite campaign data.
\citet{2022FrASS...9.8180G} re-visited FG Vir using \textit{Kepler} K2 and \textit{TESS} data and found 
around 30 significant frequencies in the K2 data, and more than 100 significant frequencies in the \textit{TESS} data.
FG Vir was observed for 52 days in 30-minute cadence photometry by the NASA \textit{Kepler} spacecraft K2 mission in 2016, 
and for 23 days in 2-minute cadence photometry by the NASA \textit{TESS} spacecraft in 2021. 
Apparently, \textit{TESS} observations of present targets are much longer than that of FG Vir.
Now, with a total of 24 sectors \textit{TESS} data, HD 52788 with 126 frequencies easily overpasses FG Vir, 
brings itself to the top tier of $\delta$ Sct-type stars with rich pulsation contents. 
This is comparable with 
the hot white dwarf pulsator PG 1159-035, in which a total of 107 frequencies were identified with 
69 days of the \textit{Kepler} satellite K2 mission 59\,s cadence data and  25 days of the \textit{TESS} satellite 20\,s cadence 
data by~\cite{2022ApJ...936..187O}. 
One can also compare it with an earlier work based on  
\textit{CoRoT} timeseries which revealed HD 50844 with hundreds of terms in the frequency range 0--30\,d$^{-1}$~\citep{2009A&A...506...85P, 2016A&A...593A..69C}. 
These authors and~\citet{2014MNRAS.439.3453B} confirmed an initial guess that $\delta$ Sct stars have a very rich frequency content. 
So HD 52788 would not be alone in frequency richness, more pulsators with dense frequencies will appear with further analyses of 
\textit{Kepler} and \textit{TESS}-like space data. 
In fact, additional examples have been published, for instance:
the binary pulsating subdwarf B star KIC 11558725 with an invisible white dwarf was revealed to have the pulsational content of 
more than 160 significant frequencies based on near-continuous 2010-2011 Kepler light curves~\citep{2012A&A...544A...1T}.
\citet{2021MNRAS.505.6166S} resolved 200+ pulsation frequencies in an oscillating Algol-type eclipsing binary system HL Dra 
with 13 \textit{TESS} sectors data (over a timespan of 352 days). 
\citet{2012ApJ...759...62B} and~\citet{2014ApJ...783...89B} revealed 349 statistically significant frequencies of the rapidly rotating $\delta$ Sct/$\gamma$ hybrid KIC 8054146 using short-cadence Kepler data during quarters Q2.3 and Q5 to Q10 which covered a time span of 27 and 557 days, respectively. 
We can further expect huge discovery opportunities of longer periodic variables provided 
by \textit{TESS}, \textit{LSST}~\citep{2019ApJ...873..111I} and other similar surveys in the near future.

\subsection{HD 53166 and HD 53349}
HD 53166 is completely a newly discovered variable star, while HD 53349 is collected in VSX and GCVS as NVS 3349 without a variability type. 
Based on their spectral types, effective temperatures and detected frequencies domain, both stars are now identified to be $\gamma$ Dor and $\delta$ Sct hybrid pulsators. 
The three strongest frequencies of both stars: 
HD 53166 at 0.8599, 2.4588, 4.9776\,d$^{-1}$;
HD 64339 at 1.888139, 1.809503,1.538036\,d$^{-1}$, 
are well in the $\gamma$ Dor pulsation domain, together with a couple of frequencies over 6\,d$^{-1}$ fitting the $\delta$ Sct pulsation. 
The above identification has followed the recent $\delta$ Sct class prototype $\delta$ Scuti (=HD 172748) advancement~\citep{2021MNRAS.505...88D}, 
$\gamma$ Dor-type descriptions of~\citet{2020MNRAS.499.3976P}, 
$\gamma$ Dor and $\delta$ Sct hybrid demonstrations of ~\citet{2020ApJ...895..124Z, 2012AJ....144..141Z, 2006CoAst.148...34R},  and 
the earlier suggestion by~\citet{2010ApJ...713L.192G}, who 
proposed a division into ’pure’ DSCT stars ($f>5$\,d$^{-1}$), ’pure’ GDOR stars ($f<5$\,d$^{-1}$), and hybrid-types exhibiting variability in both frequency regimes.

In respect to the resolved frequencies and variability types, alternative explantation could be 
that the light variation profile of HD 53166 might be a kind of rotational modulation. 
$f_0$ is very probably the rotational frequency.   
If the theoretic and effective frequency resolution 0.0009656 and 0.00152068\,d$^{-1}$ were ignored, 
$f_3$ is very closed to $f_0$ by a difference of 0.003006\cd, which would be an indication of differential rotation (spots were 
moving along the stellar surface), together with $f_4=2f_0$+0.003006\cd being possible harmonics. 
On the other hand, $f_1, 2f_1, 3f_1, 4f_1, 5f_1, 6f_1$ are very probably binarity cause with $f_1$ being the orbital frequency (corresponding orbital period of 0.40\,d). 
Thus this star could be a non-pulsating star, but a rotational variable in an eclipsing binary system of probably of 
the $\beta$ Lyr type with the depth of the secondary eclipses much smaller than those in the primary ones 
~\citep{eloy23}. 

Similarly, the third frequency term $f_2$ of HD 53349 could be regarded as a rotational frequency, 
and then $f_4=4f_2-0.003954$, $f_5=2f_2 +0.00157$. 
In Fig.~\ref{fig:6396-sp} there is another peak, nearly significant, close to $3f_2$ = 4.61\,d$^{-1}$.
An additional $\gamma$ Dor pulsation peak at $f_7$=1.2517\cd is slightly over significant level.
In resolving frequencies, we noted another peak at $f_2-0.000794$ with an equal amplitude to that of $f_2$ 
within the theoretic limit frequency resolution of 0.0009656\cd but over the adopted smallest resolvable spacing of 0.000449\cd. 
This could be an indication of rotation other than a pulsational peak.
Then, $f_2,f_4,f_5,3f_2$ are probably caused by the rotation of the star.
On the other hand, $f_0, f_1$ and $f_7$=1.2517\cd seem to be independent pulsational peaks of the GDOR type, 
regarding the temperature $T_{\rm eff}\sim$7100\,K of this star located in the region of GDOR pulsators.
The peak $f_3$=7.287296\,d$^{-1}$ seems to be an isolated pulsational DSCT peak.
In the residuals of Fig. 6, it seems there are three other peaks in the range higher than 5.0\,d$^{-1}$. 
They are probably other three independent peaks of DSCT pulsation.
So a conclusion may be drawn that HD 53349 is a hybrid $\delta$ Sct and $\gamma$ Dor pulsator 
superimposed with spots rotational variability~\citep{eloy23}. 

It is noteworthy that the above alternative rotational interpretation of light variability for the two stars HD 53166 and HD 63349 
had assumed larger errors in the resolved frequencies, which are actually at least two times the effective frequency resolution and 
ignored a frequency difference 39 times (0.003/0.000076) the calculated errors in frequencies, 
even though the possible explanation is coincidentally conform with the classification in \citet{2022arXiv221210776B}. 
The author believe in the effective frequency resolution using 24 \textit{TESS} sectors data, the frequency difference 
two times higher than its value and far beyond frequency errors cannot be ignored. 
The frequencies values and measurement uncertainties have been carefully evaluated.  
Therefore the author prefers to take the hybrid $\delta$ Sct and $\gamma$ Dor pulsators identification as final conclusion.

The matter of true intrinsic variables of these two comparison stars demonstrated the power of \textit{TESS} data.
\textit{TESS} data with 27.4 days time coverage in a sector, are especially suitable for detecting $\gamma$ Dor-type pulsations -- $\gamma$ Dor light curves on a \textit{TESS} sector are good enough for visual inspection on a computer screen. 
Previous short-term poor-quality ground observations in the literature might have misled to inaccurate resolution of light variations. This implies a clue applying \textit{TESS} data to stars with less photometry and puzzle results in the literature. 
A further search for $\gamma$ Dor stars in an extended sky region centered on HD 52788 is underway. 

\subsection{H-R Diagrams}
Taking advantage of those recently released catalogs of $\delta$ Sct stars by various all-sky surveys,
we can draw the three stars against those published known sources in the Hertzsprung-Russell diagrams for comparison.
There are quite a big volume of known $\delta$ Sct stars to compare with: 
(1) General Catalog of Variable Stars (GCVS database, Version 2022.09 V5.2,~\citealt{2017ARep...61...80S})\footnote{online URL: http://www.sai.msu.su/gcvs/cgi-bin/search.htm\#cat} designated 1018 DSCT; 267 SXPHE;
 (2) the International Variable Stars Index (VSX) database of AAVSO~\citep{AAVSO-VSX,2006SASS...25...47W}\footnote{https://www.aavso.org/vsx/} collected 27\,194; 
(3) 3939 DSCT were published in ASAS-SN Catalog of Variable Stars X~\citep{2022arXiv220502239C}\footnote{Christy et al., 2022, MNRAS, in press: https://asas-sn.osu.edu/\\variables }; ASAS-SN also listed 2231 HADS;
(4) 24\,488 contained in OGLE-IV~\citep{2021AcA....71..189S, 2020AcA....70..241P}\footnote{https://ogledb.astrouw.edu.pl/~ogle/OCVS/};  
(5) 16\,709 published in ZTF~\citep{2019PASP..131a8002B, 2019PASP..131a8003M, 2020MNRAS.499.5782O, 2020ApJS..249...18C}\footnote{Zwicky Transient Facility: https://www.ztf.caltech.edu/}; 
(6) ASAS-3 listed 3538; 
(7) Simbad archived 8440; 
(8) 14\,042 DSCT identified using TESS data by~\citet{2020MNRAS.493.5871B} and~\citet{2022arXiv221210776B}; etc. 
After removal of the duplicate stars among these catalogs, we reached a number of 70\,552 $\delta$ Sct stars from the above sources.

\begin{figure}[htb]
   \vspace{-5mm}\centering
\hspace*{-5mm}   \includegraphics[width=0.5095\textwidth, height=55mm,angle=0,scale=1.0999]{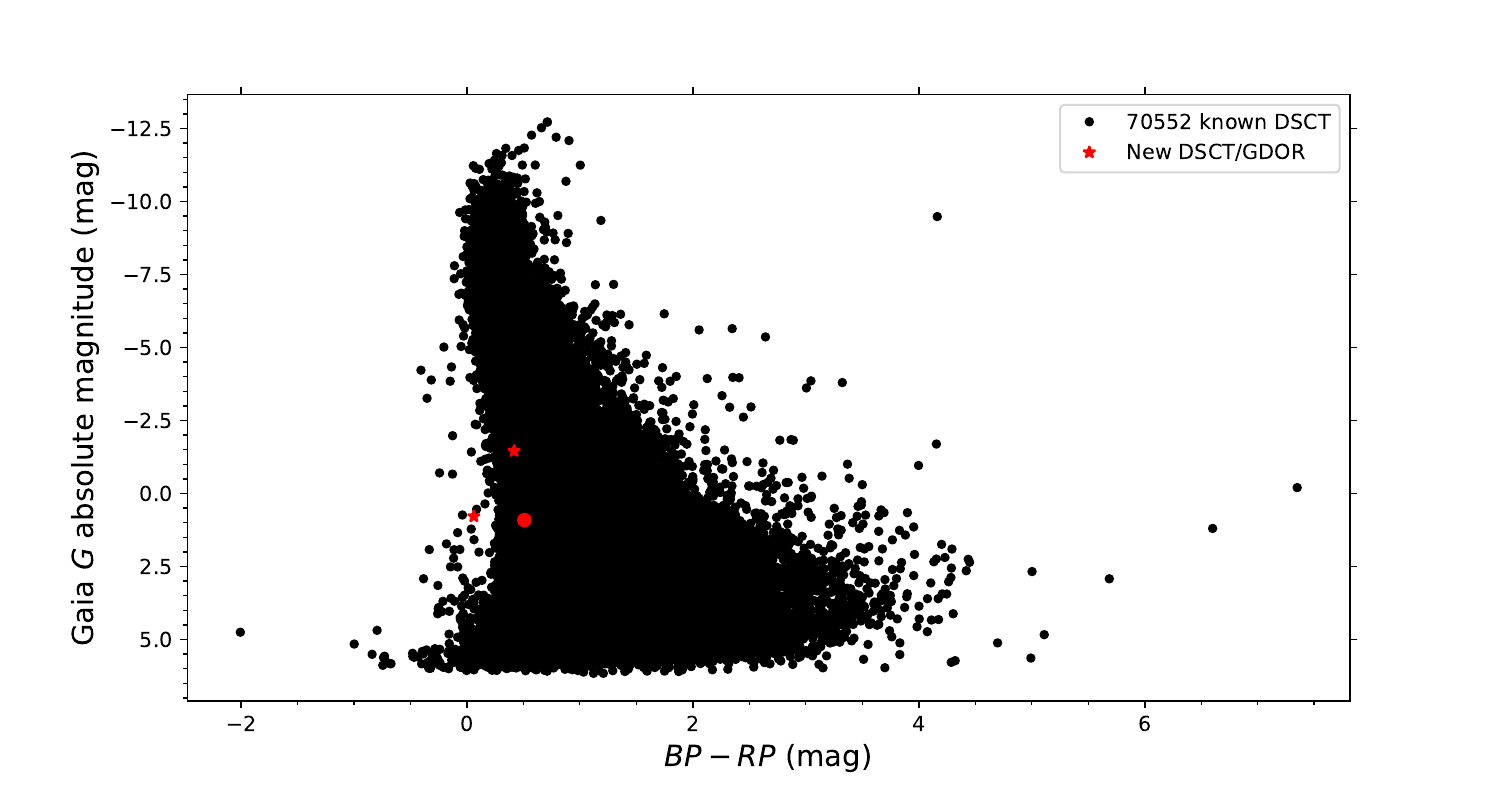}
   \caption{Present $\delta$ Sct stars in \textit{Gaia} color-magnitude diagram, plotted contrast with the known 70\,552 DSCT collected from several sources including but not limited to Simbad, GSVC, VSX, OGLE, ASAS-3, ASAS-SN, ZTF, etc. HD 52788 in red dot.}
   \label{HRD3}
\end{figure}

Using \textit{Gaia} DR2 and DR3 $G, BP,RP$ magnitude results~\citep{GaiaDR3, GaiaMission}, a \textit{Gaia} color--magnitude diagram of  $\delta$ Sct stars is drawn in Fig.~\ref{HRD3}.

\begin{figure*}[!htb]
   \vspace{-2mm}\centering
\hspace*{-26pt}   \includegraphics[width=1.02\textwidth, height=126mm,angle=0,scale=1.1299]{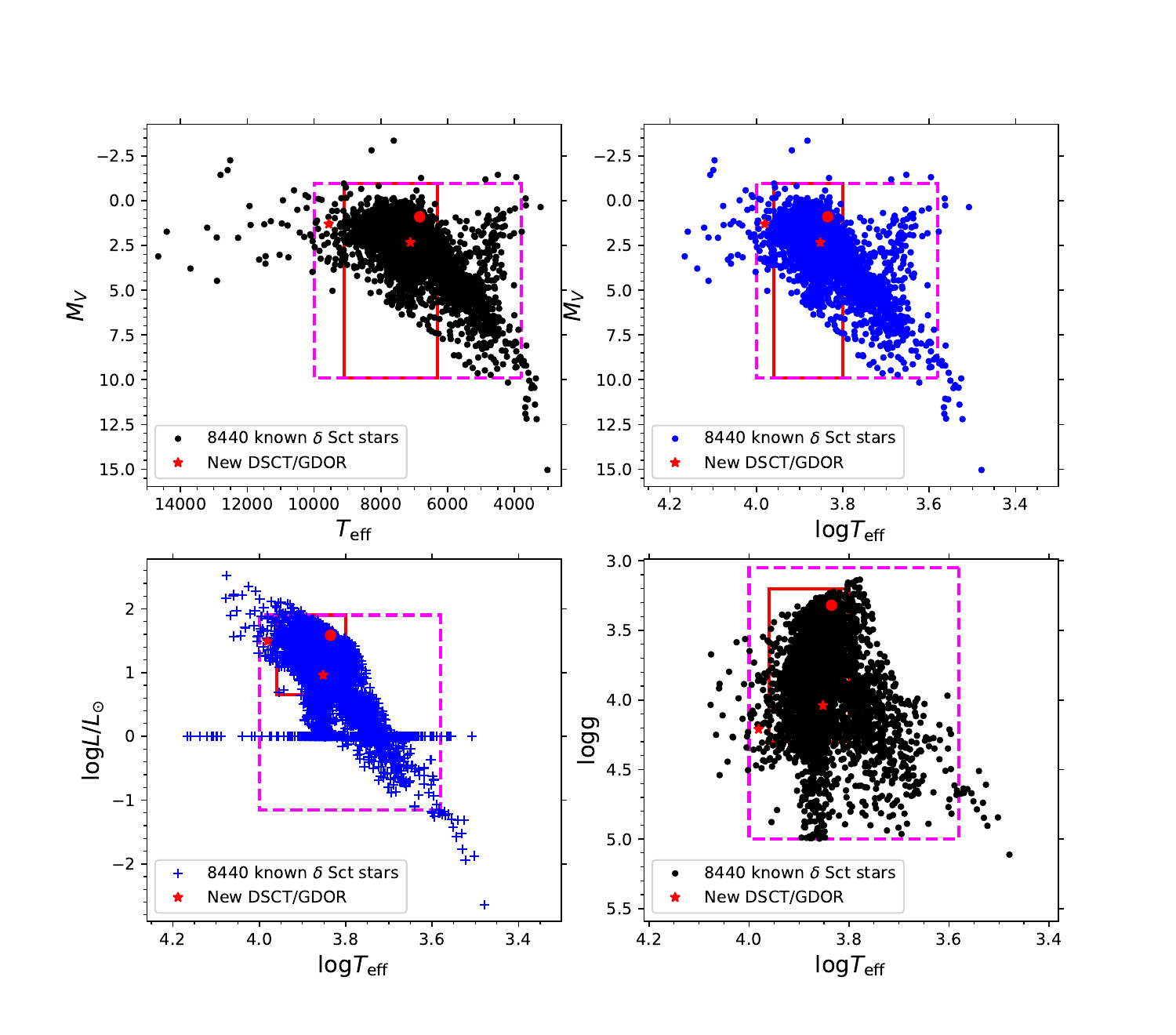}
\begin{minipage}[]{0.99\textwidth}
   \vspace{-3mm}
   \caption{Newly discovered pulsating variable stars of $\delta$ Sct (red asterisks) and HD 52788 (red dot) plotted contrast with the known 8440 DSCT extracted from Simbad. 
Red squares refer to classical $\delta$ Sct domain: $T_{\rm eff}$ in [6300, 9100], $\log T_{\rm eff}$ in [3.80, 3.96],
$\log g$ in [3.2, 4.3], $\log L/L_{\odot}$ in [0.65, 1.90] consistent with  
the H-R diagram of pulsators given by~\citet{2009AIPC.1170..403H,2008CoAst.157..240J,2000ASPC..210....3B};
magenta dashed squares refer to an extended domain: $T_{\rm eff}$ in [3800, 10000], $\log T_{\rm eff}$ in [3.58, 4.00],
$\log g$ in [3.05, 5.0], $\log L/L_{\odot}$ in [-1.15, 1.90], $M_V$ in [9.9, -0.95]. Note: $\delta$ Sct domain is explicitly much enlarged 
might be due to inaccurate stellar parameters in databases, especially lower luminosities and effective temperatures than the solar values ($\log L/L_{\odot}<0$ and $\log T_{\rm eff} <3.76, T_{\rm eff} < 5780\,K $).
}
   \label{HRD1}
\end{minipage}
\end{figure*}

\begin{figure*}[htb]
\par
\vspace*{-26mm}\centering
   \includegraphics[width=\textwidth, height=68mm,angle=0,scale=1.019]{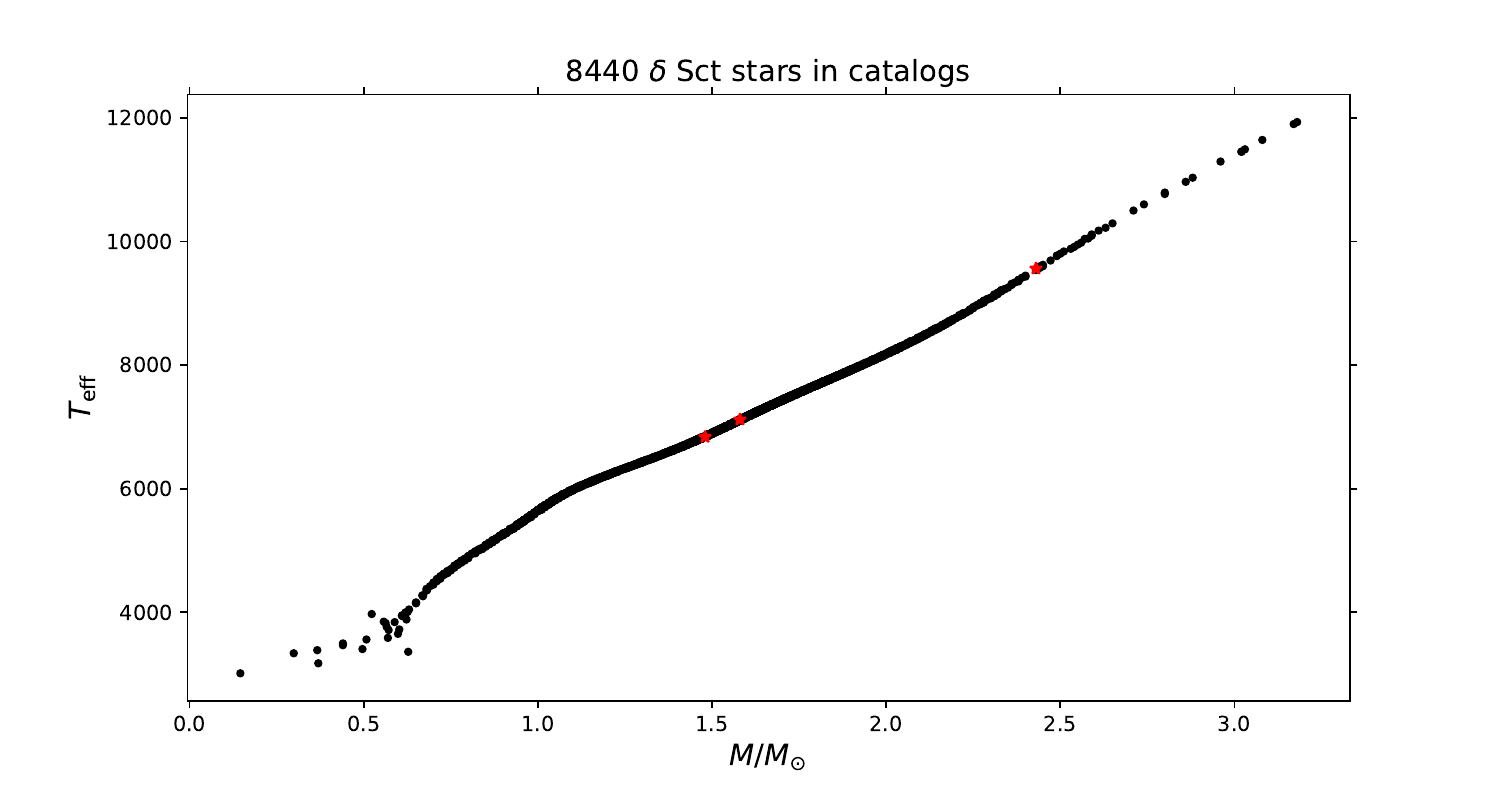}
\begin{minipage}[]{0.98\textwidth}
   \caption{Present $\delta$ Sct stars (red asterisks) plotted contrast with the known 8440 DSCT extracted from Simbad.}
   \label{HRD2}
\end{minipage}
\end{figure*}

We further extracted effective temperature, luminosity, surface gravity, parallax, distance and other stellar parameters from 
\textit{TESS} Input Catalog~\citep[TIC v8.2,][]{TICv82} and \textit{Gaia} DR2/DR3 for the collection.
However, a good number of $\delta$ stars from the above surveys have less luminosities than the Sun, 
we instead tend to draw additional H-R diagrams only for those $\delta$ Sct stars collected in Simbad 
rather than VizieR.
It is known that Simbad/CDS provides information on astronomical objects which have been studied in scientific articles, 
while the CDS VizieR database is a complementary research tool which is composed of various survey catalogs.
Figures~\ref{HRD1}--\ref{HRD2} show the three stars compared with 8440 $\delta$ Sct stars extracted from Simbad (as of September 2022).
As shown, even only the Simbad 8440 $\delta$ Sct stars, there are still considerable number with lower luminosities and effective temperatures than solar values. 
If all collected DSCT were used, the H-R diagrams would show a much extended DSCT domain than 
the classical borders that demonstrated in the literature
\citep[see fig.1 in][]{2009AIPC.1170..403H, 2008CoAst.157..240J, 2001A&A...366..178R,2000ASPC..210....3B} and 
in the recent studies of a sample of 15\,000+ \textit{Kepler} A and F stars 
in and around the $\delta$ Scuti instability strip using \textit{Gaia} DR2 parallaxes by~\citet{2019MNRAS.485.2380M}. 

\begin{figure*}[!htb]
   \vspace{-12mm}\centering
\hspace*{-26pt}   \includegraphics[width=1.02\textwidth, height=126mm,angle=0,scale=1.1299]{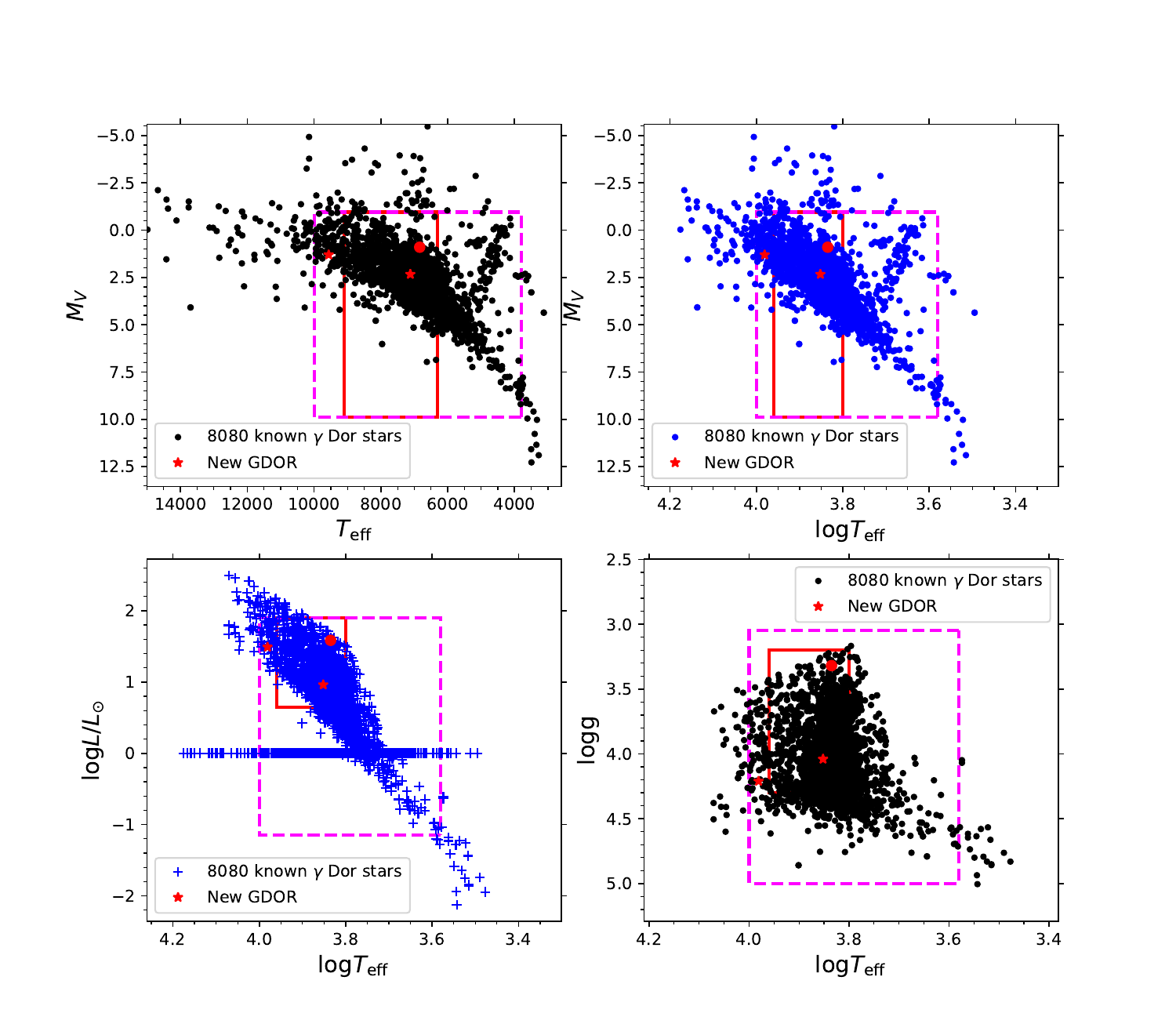}
\begin{minipage}[]{0.99\textwidth}
   \vspace{-3mm}
   \caption{Newly discovered pulsating variable stars of $\gamma$ Dor--$\delta$ Sct hybrid (red asterisks) and HD 52788 (red dot) plotted contrast with the known 8080 GDOR. The same squares as Fig.~\ref{HRD1}. Note: $\gamma$ Dor domain is explicitly much enlarged 
might be due to inaccurate stellar parameters in databases, especially lower luminosities and effective temperatures than the solar values ($\log L/L_{\odot}<0$ and $\log T_{\rm eff} <3.76, T_{\rm eff} < 5780\,K $).
}
   \label{HRD4}
\end{minipage}
\end{figure*}

Similarly, we have collected $\gamma$ Dor stars from multiple sources:
(1) dozens from earlier pioneering works~\citep{2007AJ....133.1421H, 2005AJ....129.2815H, 1999PASP..111..840K};
(2) 120+ from~\citet{2022AJ....163..180H, 2011AJ....142...39H};
(3) 611 and 207 resulted from \textit{Kepler} data by~\cite{2020MNRAS.491.3586L} and \citet{2015AJ....149...68B},respectively;
(4) 108 in GCVS v5.2 (as of 2022 Oct.19); 
(5) 1032 in VSX-AAVSO (as of 2022 Nov.15);
(6) 1641 in Simbad (as of 2022 Nov.15); 
(7) 6358 based on TESS data identified by~\citet{2022arXiv221210776B} and~\citet{2020MNRAS.493.5871B}; etc.
After removal of the duplicate stars among these sources, we reached a number of 8\,080 
$\gamma$ Dor stars with stellar parameters extracted from \textit{Gaia} DR2/DR3 and TIC v8.2. 
A H-R diagram of $\gamma$ Dor stars is then drawn in Fig.~\ref{HRD4}, 
where 8080 GDOR from the above sources were used.

It is noted that both DSCT and GDOR domains are much extended to lower luminous region. This bias with classical ranges 
probably be resulted either from inaccurate parameters or false memberships of each type stars. 
In this regard, a further examination and confirmation of bona fide members of each one of DSCT and GDOR 
in the mixed collections is urgently needed. 
For instance, \citet{2015AJ....149...68B} found 207 $\gamma$ Dor candidates and 84 $\delta$ Sct candiates from \textit{Kepler} data. 
These candidates could have been taken as true members in some relevant catalogs -- a duplication check shows 125 of 207 
being collected in other publications.   
The star BS 8799 (=HD 218396 =TIC 245368902=V342 Peg) in \citet{1999PASP..111..840K}'s confirmed list of GDOR, its variability type is controversial 
as Simbad labels it to be an ellipsoidal variable other than GDOR. However, it is a genuine GDOR, \textit{TESS} data on sector 56 (2022.09.02) show clearly typical $\gamma$ Dor-type light variations.   
Those discoveries resulted from automatic identification using machine learning methods based on large sky surveys are huge and 
would have mixed with some non-members. The author is thus inspired to examine and verify the candidates of various sources with \textit{TESS} data in a up-coming work. 

On the other hand, \citet{2020MNRAS.493.5871B} and~\citet{2022MNRAS.510.5743B} who have carefully chosen and calibrated effective temperature in sample stars 
to ensure a safe variability classification.
The spectral type directly related to temperature is important for ascertaining the variability class. 
Rigorous observational H-R diagrams for both $\delta$ Sct and $\gamma$ Dor bona fide members based on reliable temperature and luminosity 
by taking account of \textit{Gaia} DR3 parameters corrected for interstellar extinction would be much persuasive and meaningful.
With such convincing H-R diagrams we can compare them with theoretic borders of instability strip connecting to oscillation mechanisms of $\delta$ Sct and $\gamma$ Dor stars such as that by \citet{2005A&A...435..927D} and \citet{2016MNRAS.457.3163X}. 

The author would like to 
conduct an asteroseismology for these three stars upon making relevant stellar evolution and oscillation modeling programs available to run in the near future. 
At present we report the full analysis of light curves and the results of 
new discoveries of pulsational variability of HD 53166 and HD 53349, 
and a complete update of HD 52788 light variations complexity.  
%

%

\vspace{03.0cm}
{\it This manuscript was prepared following `elsarticle-template-harv.tex' }

\section*{Acknowledgements}
I appreciate the two anonymous referees for their thoughtful comments and meticulous reading 
which allowed the author to greatly improve the quality of the manuscript. 
I am indebted to my wife for her unwavering support throughout my research.
This work could not have been done without my wife’s continuous support. 
This paper includes data collected with the \textit{TESS} mission. 
Funding for the \textit{TESS} mission is provided by the NASA Explorer Program. 
\textit{TESS} data were obtained from the MAST data archive at the Space Telescope Science Institute (STScI). 
STScI is operated by the Association of Universities for Research in Astronomy, Inc., under NASA contract NAS 5–26555.
We acknowledge the use of \textit{TESS} data, which are derived from
pipelines at the \textit{TESS} Science Processing Operations Center (SPOC). 
\textit{TESS} High Level Science Products (HLSP) produced by the Quick-Look Pipeline (QLP) at 
the \textit{TESS} Science Office at MIT, which are publicly available from the Mikulski Archive for Space Telescopes (MAST). 
This research has made use of the SIMBAD/VizieR databases, operated at CDS, Strasbourg, France.
This research has made use of the International Variable Star Index (VSX) database, operated at AAVSO, Cambridge, Massachusetts, USA.
Special acknowledgements go to all the projects and surveys mentioned in this work including but not limited to: 
OGLE-V, ASAS-3, ASAS-SN, and ZTF.
This work has made use of data from the European Space Agency (ESA) mission
{\it Gaia} ({https://www.cosmos.esa.int/gaia}), processed by the {\it Gaia}
Data Processing and Analysis Consortium (DPAC\footnote{https://www.cosmos.esa.int/web/gaia/dpac/consortium}). Funding for the DPAC
has been provided by national institutions, in particular the institutions
participating in the {\it Gaia} Multilateral Agreement.
This work made use of \textit{Astropy}:\footnote{http://www.astropy.org} a community-developed core Python package and an ecosystem of tools and resources for astronomy \citep{astropy:2013, astropy:2018, astropy:2022}; 
\textit{Astroquery}~\citep{astroquery:2019};
\textit{Lightkurve}~\citep{lightkurve}; 
\textit{Matplotlib}~\citep{Hunter:2007};
\textit{SciPy}~\citep{SciPy2020};
\textit{NumPy}~\citep{NumPy2020}.
\appendix

\section{A Catalog of 70552 $\delta$ Scuti Stars}

Based on $\delta$ Sct stars listed in several existing catalogs and databases by Simbad, OGLE, VSX, ASAS-SN, ZTF, GCVS, 
and other sources including three dozen sporadic new DSCTs picked up from recent publications (e.g. \citealt{2022arXiv220811915K, 2021PASP..133e4201S}) and 14042 DSCT from \textit{TESS} data by~\citet{2020MNRAS.493.5871B, 2022arXiv221210776B}, 
we reached 
an up-to-date comprehensive catalog of $\delta$ Sct stars that consists of 70\,552 individual stars after removing those duplicated entires.
For those 14042 DSCT, the main identifiers if available in Simbad were extracted and compiled in the list. 
In fact, there are 9091, 706 and 1559 with HD, BD and TYC identifiers, respectively.

Each entry in the catalog is comprised of 26 columns: 
TIC, source ID or main ID of Simbad, \textit{Gaia} DR3 ID, RA\_Dec/2MASS\_ID, $B$\_TIC, $V$\_TIC,
$T_{\rm eff}$\_TIC, $T_{\rm eff}$\_Gaia,  $\log g$\_TIC, $\log g$\_Gaia,
Mass\_TIC, $R$\_TIC, $R$\_Gaia, $L$\_TIC, $L$\_Gaia, parallax(mas), distance(pc), $M_V$,
$\log T_{\rm eff}$\_TIC, $\log L$\_TIC, $G$\_mag, $BP$\_mag, $RP$\_mag, $BP-RP$, $BP-G$, $G-RP$, Radial\_Velocity\_Gaia and errors for a few parameters.

The whole catalog in its entirety is provided as online materials in both machine-readable text format (CSV file) 
and human-friendly HTML version\footnote{https://deltascuti.wixsite.com/delta/catalog}.
Interested readers may enquire it from the author via email or 
download full table in CSV file on the \href{https://ars.els-cdn.com/content/image/1-s2.0-S1384107623000829-mmc1.csv}{journal's web} or \href{https://deltascuti.wixsite.com/delta/download}{\url{https://deltascuti.wixsite.com/delta/download}} . 
A reduced demo is given in Table~\ref{Catalog-DSCT}. 
More importantly, in the HTML version, each TIC ID is hyperlinked to STScI archive MAST in a default radius of 0.00033$^\circ$ of the target,
each \textit{Gaia} DR3 ID with hyperlink to \textit{Gaia} data at VizieR/CDS,
and main IDs or coordinates are linked to CDS Portal.
The HTML version makes consulting, checking, follow-up and downloading \textit{TESS}/\textit{Kepler}/K2 and \textit{Gaia} DR3 data unprecedentedly convenient and efficient.


\setcounter{table}{0}
\begin{sidewaystable} 
\footnotesize 
\scriptsize
\setlength{\tabcolsep}{3pt}  
\renewcommand{\arraystretch}{1.2}  
\vspace*{-80mm}
  \caption[]{A combined catalog of 70,552 $\delta$ Sct ctars (reduced demo version).  Full table in CSV file is on the  \href{https://ars.els-cdn.com/content/image/1-s2.0-S1384107623000829-mmc1.csv}{journal's web} or \href{https://deltascuti.wixsite.com/delta/download}{\url{https://deltascuti.wixsite.com/delta/download}} .}
  \label{Catalog-DSCT}
  \begin{center}
  \begin{tabular}{lrlrrrrrcrr}
  \hline\noalign{\smallskip} 
\multicolumn{1}{c}{TIC\_ID} &   \multicolumn{1}{c}{           Source ID}&\multicolumn{1}{c}{              Gaia DR3}&\multicolumn{1}{c}{             2MASS/RA\_Dec} &   
$B$\_TIC  &    $V$\_TIC  & $T_{\rm eff}$\_TIC  &  $T_{\rm eff}$\_Gaia  &  $\log g$\_TIC  & $\log g$\_Gaia  
& $\cdots\cdots$ \\
  \hline
TIC 1000089523      & ZTFJ141934.31+113547.9& 1225642756991957120& 214.892981 11.596656 &
      nan&  18.6731&      nan&  7297.78&      nan&   4.3895& $\cdots\cdots$ \\
TIC 1000101652      & ZTFJ140848.41+123606.6& 1226608196920814848&  212.201759 12.601821 &      nan&  18.4599&   7019.0&  7150.74&    nan&  4.5852& $\cdots\cdots$ \\
TIC 1000104768      & ZTFJ141805.41+131126.6& 1226865933613176448&  214.522574 13.19073 &      nan&  17.5211&   6977.0&  6847.88&  4.80123&    4.469& $\cdots\cdots$ \\
TIC 100012164       & TIC 100012164& 864002769753103616&  173.268106 63.89601 &    9.552&    9.253&   7376.0&  7211.87&  4.09371&   3.9797&  $\cdots\cdots$ \\
TIC 1000156752      & ZTFJ135719.50+152149.3& 1231155948812038016&  209.331291 15.363725 &      nan&  18.7852&      nan&  7277.38&      nan&   4.3927& $\cdots\cdots$ \\
TIC 1000164869      & ZTFJ141730.31+150946.1& 1231876678683202816&  214.376322 15.162809 &      nan&   18.897&      nan&  7405.75&      nan&   4.4463&  $\cdots\cdots$ \\
\multicolumn{1}{c}{$\vdots$}\\
TIC 99968017        & ZTFJ192910.74-071936.8& 4207666832973112960&  292.294793 -7.326884 &   15.583&   15.469&   6987.0&  7740.32&  4.10928&   4.1564&$\cdots\cdots$\\
TIC 99971924        & ZTFJ112629.01+612732.7& 862543889621373696&  171.621022 61.459218 &   14.933&   15.045&   7537.0&  7441.75&  4.36968&   3.9848&$\cdots\cdots$\\
TIC 99978409        & ZTFJ204145.78+334207.8& 1863327001102874240&  310.440792 33.702168 &   15.369&   14.673&   6938.0&  9015.28&  3.30249&   3.5687&$\cdots\cdots$\\
TIC 99983100        & ASASSN-V J204212.17+323517.2& 1863120877024811264&  20 42 12.17 +32 35 17.2 &      nan&  15.7216&   5886.0&  5831.14&  3.32024&4.0683&$\cdots\cdots$\\
TIC 99995027        & HD 175355& 4310238314231510528&  283.67868 8.099308  &    9.366&    9.142&   8111.0&  8395.73&  4.26642&   4.2661& $\cdots\cdots$\\
  \hline
\end{tabular}
\bigskip  
\vspace{15.5mm}
  \caption[]{A combined catalog of 8080 $\gamma$ Dor stars (reduced demo version). Full table in CSV file is on the \href{https://ars.els-cdn.com/content/image/1-s2.0-S1384107623000829-mmc2.csv}{journal's web} or \href{https://deltascuti.wixsite.com/delta/download}{\url{https://deltascuti.wixsite.com/delta/download}} . }
  \label{Catalog-GDOR}  
  \vspace{3.5mm}
  \begin{tabular}{lrlrrrrrcrrrrr}
  \hline\noalign{\smallskip}
\multicolumn{1}{c}{TIC\_ID} &\multicolumn{1}{c}{              Source ID}& \multicolumn{1}{c}{             Gaia DR3}&\multicolumn{1}{c}{             2MASS/RA\_Dec} &   
$B$\_TIC  &    $V$\_TIC  & $T_{\rm eff}$\_TIC  &  $T_{\rm eff}$\_Gaia  &  $\log g$\_TIC  & $\log g$\_Gaia  & 
Mass\_TIC& $\cdots\cdots$ \\
  \hline
TIC 100008712       & TIC 100008712& 863683499063563648&  J112953.36+630613.9 &   10.129&     9.83&   7080.0&  6946.84&  4.27086&   4.1372&   1.57& $\cdots\cdots$ \\
TIC 100102390       & HD 10167& 4958910970141833216&  J013830.73-425540.6 &    6.984&     6.67&  7008.68&      nan&   3.8645&      nan&   1.54& $\cdots\cdots$ \\
TIC 1001512783      & TIC 153989642& 1685166397526839424&  --                  &      nan&   8.7316&   6308.0&  6187.54&      nan&   4.2671&    nan&$\cdots\cdots$ \\ 
TIC 1002938519      & TIC 379081616& 3726490479878642688&  --                  &      nan&   5.5506&      nan&      nan&      nan&      nan&    nan& $\cdots\cdots$\\
TIC 100325374       & TIC 100325374& 6668210166134920192&  J202832.56-484039.2 &     9.62&    9.242&   7070.0&  6913.17&  4.11936&   4.0608&   1.56&$\cdots\cdots$\\
TIC 100378749       & TIC 100378749& 6474064587655163776&  J203049.99-525234.2 &    9.489&    9.077&   6660.0&   6554.4&  4.20033&   4.1194&    1.4&$\cdots\cdots$\\
\multicolumn{1}{c}{$\vdots$}\\
TIC 99842485        & TIC 99842485& 1054696912513345280&  J110332.30+625725.2 &    9.705&    9.293&   6376.0&      nan&      nan&      nan&    nan&$\cdots\cdots$\\
TIC 99867019        & TIC 99867019& 861500899763587456&  J110535.22+602347.8 &   10.144&    9.688&   6488.0&      nan&      nan&      nan&    nan&$\cdots\cdots$\\
TIC 99883676        & TIC 99883676& 4200367450155720960&  J192812.96-093216.8 &    8.679&     8.38&   7130.0&  7023.34&   4.2454&   4.1323&  1.587&$\cdots\cdots$\\
TIC 99898937        & TIC 99898937& 862289764996484096&  J110913.24+622951.8 &    9.537&    9.259&   7477.0&  7408.35&  4.19653&   4.1017&   1.72&$\cdots\cdots$\\
TIC 99971569        & TIC 99971569& 863973357816605440&  J112609.67+634936.5 &    9.278&    8.968&   7158.0&  6999.44&  3.93968&   3.9323&    1.6&$\cdots\cdots$\\
  \hline
\end{tabular}
\end{center}
\end{sidewaystable}\normalsize
\newpage
 
\section{A Catalog of 8080 $\gamma$ Doradus Stars}
We have compiled a catalog of $\gamma$ Dor stars based on the sources of 
GCVS, VSX, \textit{Kepler} data~\citep{2020MNRAS.491.3586L},~\citet{2022AJ....163..180H, 2011AJ....142...39H, 2007AJ....133.1421H}, Simbad, a few thousand from \textit{TESS}~\citep{2020MNRAS.493.5871B, 2022arXiv221210776B} and others. 
After removal of the duplicate stars among these sources, 
we reached a number of 8080 individual $\gamma$ Dor stars with stellar parameters extracted from \textit{Gaia} DR2/DR3 and TIC v8.2.  Parameters or columns are the same as the above $\delta$ Sct stars catalog. 
The whole catalog in its entirety is provided as online materials in a machine-readable text format (CSV file). 
Interested readers may request it from the author via email or 
download on the \href{https://ars.els-cdn.com/content/image/1-s2.0-S1384107623000829-mmc2.csv}{journal's web} or \href{https://deltascuti.wixsite.com/delta/download}{\url{https://deltascuti.wixsite.com/delta/download}}.  
A reduced demo is given in Table~\ref{Catalog-GDOR}. 


\clearpage
\bfr{Note: References with 5+ authors may be cutoff and replaced by `et al.'  in this manuscript.} 
\bibliographystyle{elsarticle-harv} 
\bibliography{../raa/zayRef}




\end{document}